\documentclass[aps,pra,9pt,twocolumn,showpacs,superscriptaddress]{revtex4-2}
\usepackage{ulem}
\usepackage{physics}							
\usepackage{latexsym}
\usepackage{amssymb}
\usepackage{bm}
\usepackage{graphics,epstopdf}
\usepackage{color}
\usepackage{newlfont}
\usepackage{amsfonts}
\usepackage{amsthm}
\usepackage{graphicx}
\usepackage{epsfig}
\usepackage[colorlinks=true,linkcolor=blue,citecolor=red,plainpages=false,pdfpagelabels]{hyperref}
\usepackage{amsmath}

\DeclareOldFontCommand{\rm}{\normalfont\rmfamily}{\mathrm}

\usepackage{times}

\newcommand{\be}{\begin{equation}}
\newcommand{\ee}{\end{equation}}
\newcommand{\bc}{\begin{center}}
\newcommand{\ec}{\end{center}}
\newcommand{\bea}{\begin{eqnarray}}
\newcommand{\eea}{\end{eqnarray}}
\newcommand{\ba}{\begin{array}}
\newcommand{\ea}{\end{array}}

\begin{document}
\title{Generalised quantum speed limit for arbitrary time-continuous evolution}
\author{Dimpi Thakuria}\email{dimpithakuria97@gmail.com}

\affiliation{\(\)Harish-Chandra Research Institute,  A CI of Homi Bhabha National Institute, Chhatnag Road, Jhunsi, Prayagraj  211019, Uttar Pradesh, India
}
\affiliation{\(\)Atominstitut, Technische Universität Wien, Stadionallee 2, 1020 Vienna, Austria
}
\author{Abhay Srivastav}\email{abhaysrivastav@hri.res.in}
\affiliation{\(\)Harish-Chandra Research Institute,  A CI of Homi Bhabha National Institute, Chhatnag Road, Jhunsi, Prayagraj  211019, Uttar Pradesh, India
}
\author{Brij Mohan}\email{brijhcu@gmail.com}
\affiliation{\(\)Harish-Chandra Research Institute,  A CI of Homi Bhabha National Institute, Chhatnag Road, Jhunsi, Prayagraj  211019, Uttar Pradesh, India
}
\affiliation{Department of Physical Sciences, Indian Institute of Science Education and Research (IISER), Mohali-140306, India}

\author{Asmita Kumari}\email{asmitakumari@hri.res.in}
\affiliation{\(\)Harish-Chandra Research Institute,  A CI of Homi Bhabha National Institute, Chhatnag Road, Jhunsi, Prayagraj  211019, Uttar Pradesh, India
}
\author{Arun Kumar Pati}\email{akpati@iiit.ac.in}
\affiliation{\(\)Centre for Quantum Science and Technology (CQST), International Institute of Information Technology, Hyderabad, Gachibowli, Telangana 500032, India}

\begin{abstract}
The quantum speed limit describes how quickly a quantum system can evolve in time from an initial state to a final state under a given dynamics. Here, we derive a generalised quantum speed limit (GQSL) for arbitrary time-continuous evolution using the geometrical approach of quantum mechanics. The GQSL is applicable for quantum systems undergoing unitary, non-unitary, completely positive, non-completely positive and relativistic quantum dynamics. This reduces to the well known standard quantum speed limit (QSL), i.e., the Mandelstam-Tamm bound when the quantum system undergoes unitary time evolution. Using our formalism, we then obtain a quantum speed limit for non-Hermitian quantum systems. To illustrate our findings, we have estimated the quantum speed limit for a time-independent non-Hermitian system as well as for a time-dependent non-Hermitian system namely the Bethe-Lamb Hamiltonian for general two-level system.
\end{abstract}

\maketitle
\section{Introduction}

The laws of quantum physics impose a fundamental limit on the evolution speed of quantum systems. This limit is commonly known as the quantum speed limit (QSL). The QSL establishes a lower bound on the evolution time required to transport a quantum system from an initial state to the target state under a given dynamics. This bound provides a novel relation between energy and time, like the energy-time uncertainty relation. The study of QSL is essential not only for theoretical understanding, but also for practical applications in the rapidly developing field of quantum technologies. In recent years, QSL has been extensively investigated for applications in quantum computing~\cite{AGN12}, quantum batteries~\cite{F.Campaioli2018,Manab2020, Mohan2021,Mohan21}, quantum control theory~\cite{Caneva2009,Campbell2017}, quantum metrology~\cite{Demkowicz2012,Campo2013,Toth2014,Beau2017,Campbell2018}, quantum thermodynamics~\cite{Campo2014,Mukhopadhyay2018} etc.

The first QSL bound on the evolution time was discovered by Mandelstam and Tamm~\cite{Mandelstam1945} using the Heisenberg-Robertson uncertainty relation. It depends on the shortest distance between the initial and the final state and the variance of the driving Hamiltonian. Later on, another QSL bound was given by Margolous and Levitin~\cite{Margolus1998}, which depends on the mean energy of the driving Hamiltonian. Although Aharonov and Anandan first introduced the notion of the speed of transportation of a quantum system using the Fubini-Study metric, which is now known as the `quantum speed limit'~\cite{Anandan1990}. Subsequently, the same notion was defined in Ref.~\cite{akp91}, where the Riemannian metric was used to define the speed of transportation of quantum systems for unitary evolutions. It was shown that the speed of transportation of a quantum state in the projective Hilbert
space is proportional to the variance of the driving Hamiltonian of the quantum
system. The quantum speed of transportation was discovered to be proportional to the super current in the Josephson junction~\cite{Anandan1997}. The notion of the Fubini-Study metric has been generalised for arbitrary quantum evolution in Ref.~\cite{Pati1995}.

The dynamics of a closed quantum system is unitary. However, it no longer remains unitary when the system interacts with its surroundings. In this setting, the dynamics can be described by a Liouvillian, particularly by a Lindblad master equation, if the system's time evolution is Markovian and completely positive trace preserving (CPTP). However, in some cases, the effective dynamics of such a system is governed by a non-Hermitian Hamiltonian. If a quantum system exchanges energy, particles, etc., with its surroundings, then the effective dynamics of such a quantum system is described by a non-Hermitian Hamiltonian that generates gain and loss~\cite{Roccati2021,Roccati2022}. If gain and loss are balanced, then the Hamiltonian becomes $PT$-symmetric \cite{Ali2002,bender2002,bender2003}. Recently, quantum systems with non-Hermitian Hamiltonians have been experimentally realized in trapped ions~\cite{Ding2021,Wang2021,Ding2022}, cold atoms~\cite{Li2019,Jiang2019}, nitrogen-vacancy centers~\cite{Liu2021,Wu2019}, and superconducting circuits~\cite{Part2019}. In general, the dynamics of a quantum system can be arbitrary, which can be unitary or non-unitary, driven or undriven, linear or nonlinear, relativistic or non-relativistic. Moreover, the dynamics of the system can be completely positive or non-completely positive also. In the rest of the paper, we refer all such dynamics as arbitrary time-continuous dynamics.

The concept of quantum speed limit has been widely studied for closed quantum system dynamics~\cite{Mandelstam1945,Anandan1990, Margolus1998,Levitin2009, Gislason1956, Eberly1973, Bauer1978, Bhattacharyya1983, Leubner1985,  Vaidman1992, Uhlmann1992, Uffink1993, Pfeifer1995,  Horesh1998, AKPati1999, Soderholm1999,  Andrecut2004, Gray2005, Luo2005,  Zielinski2006,  Andrews2007, Yurtsever2010, Fu2010, Zwierz2012, Poggi2013,Kupferman2008, Jones2010, Chau2010, S.Deffner2013, Fung2014, Andersson2014, D.Mondal2016, Mondal2016, S.Deffner2017, Campaioli2018,Giovannetti2004, Batle2005, Borras2006, Zander2007,Ness2022,Bagchi2022,Pati2023} as well as for open quantum system dynamics~\cite{Taddei2013,Campo2013,Deffner2013, Fung2013, Pires2016, S.Deffner2020,Jing2016,Funo2019}. It is essential to mention that a method to measure speed of quantum system in interferometry was proposed in Ref.~\cite{D.Mondal2016} and recently, an experiment was reported where the quantum speed limits were tested in a multi-level quantum system by tracking the motion of a single atom in an optical trap using fast matter-wave interferometry~\cite{Ness}. One question that remains open until now is that is there a quantum speed limit bound that applies to arbitrary time-continuous dynamics? The answer to this question can have a wide range of applications in quantum physics, quantum information, quantum computing and quantum  control theory. Therefore, the study of QSL for arbitrary dynamics of a quantum system is of prime importance.

In this paper, by employing the geometrical approach of quantum mechanics, we derive the quantum speed limit for a quantum system undergoing arbitrary time-continuous dynamical process. The proof of the generalised QSL needs the notion of the most general definition of the Fubini-Study metric defined in Ref.~\cite{Pati1995}. This speed limit depicts a fundamental limitation on the minimal time required for the quantum system to evolve from a given initial state to a final state undergoing arbitrary dynamical processes. Our generalized QSL reduces to the well known Mandelstam-Tamm bound for unitary evolution of quantum system. Using our formalism then, we obtain a QSL bound for non-Hermitian quantum systems. By utilizing these findings, we explore the quantum speed limit for a time-independent non-Hermitian system with gain and loss and a time-dependent non-Hermitian system namely the Bethe-Lamb Hamiltonian for general two-level system. Our result can also be applied to $\mathcal{PT}$-symmetric systems. Thus, the generalised QSL proved here can have a wide variety of applications hitherto unexplored.

Our work is organized as follows. In Section-\ref{QSL:arbitaray}, we provide the geometrical derivation of the generalised quantum speed limit (GQSL) for arbitrary time-continuous evolutions of pure as well as mixed states. In Section-\ref{QSL:non-Hermition}, we derive the generalised quantum speed limit for non-Hermitian systems. In Section-\ref{QSL:gain-loss}, we investigate the GQSL for a non-Hermitian two-level quantum system where gain and loss are present. In Section-\ref{QSL:BH}, we estimate the GQSL for the Bethe-Lamb Hamiltonian of the two-level quantum system. Finally, in the last Section, we provide the conclusions.

 \section{QSL for arbitrary quantum evolution}\label{QSL:arbitaray}
 
 Consider a quantum system with an associated complex separable Hilbert space ${\cal H}^N$, where dim$\cal{H} = N$.
 A pure quantum state $\ket{\Psi} \in \cal{H}$ can represent the state of the system. 
  In quantum theory, pure quantum states are described by equivalence classes $\{\Psi\}$ of unit vectors $\ket{\Psi} \in {\cal H}$, where two unit vectors are equivalent if they differ by a 
 phase factor.  The set of the equivalence classes $\{\Psi\}$ under the projection map is called the projective Hilbert space 
 ${\cal P}({\cal H})$. Thus, the elements of the projective Hilbert space ${\cal P}({\cal H})$ are equivalence classes $\{ \Psi \}$ of vectors of ${\cal H}$ and $\ket{\Psi}$ need not be a unit vector. More precisely, we say that any two vectors of ${\cal H}^* := {\cal H} - \{0\}$ are equivalent if they differ by any complex factor, and define ${\cal P}({\cal H})$ to be the set of the corresponding equivalence classes. 
 When $\ket{\Psi(t) }$ evolves in time it traces a curve in ${\cal H}$, i.e. $t: \rightarrow  \ket{\Psi(t) }$, $0 \le  t \le T$ is a
curve on the Hilbert space. Note that here, the evolution is not necessarily unitary. It can be unitary, non-unitary, completely positive, not-completely positive and even beyond the Schr{\"o}dinger evolution. If the evolution is unitary, then the time evolution of the quantum system is generated by the one-parameter group $e^{-\frac{iHt}{\hbar}}$ of unitary operators where $t\in \mathbb{R}^+$ and $H$ is the self-adjoint Hamiltonian operator acting on $\cal{H}$. This evolution curve in ${\cal H}$ may project to either a closed or an open curve in the projective
Hilbert space ${\cal P}({\cal H}) $  depending on the type of evolution, i.e., whether the evolution is 
cyclic or non-cyclic. 
 
We now derive the quantum speed limit for pure as well as for mixed states of quantum systems undergoing arbitrary time-continuous evolution.

 \subsection{For pure initial state}
 The projective Hilbert space  ${\cal P}({\cal H})$ of quantum systems undergoing arbitrary time-continuous evolution is equipped with a natural metric which can be defined using the Fubini-Study metric \cite{Pati1995}.
 The generalized Fubini-Study distance for any two unormalised vectors $\ket{\Psi_1} \in {\cal H} $ and $\ket{\Psi_2} \in {\cal H} $ can be defined as 
 \begin{equation}
 S^2 = 4 \left(1 - \left|\frac{ \langle \Psi_2| }{\norm{\Psi_2 }} \frac{ |\Psi_1\rangle}{\norm{\Psi_1 }} \right|^2 \right).
 \end{equation}

  The above distance satisfies identity, symmetry and triangle inequalities.  This definition of the Fubini-Study metric is a generalisation of the one given in Ref.~\cite{Anandan1990}. The Fubini-Study metric defined by Anandan and Aharanov in Ref.\cite{Anandan1990} is invariant under $U(1)$ transformations only, where the vectors are multiplied by some phase factors. However, the generalized definition of the Fubini-Study metric is invariant under more general transformations of the kind $\ket{\Psi_i} 
  \rightarrow Z_i \ket{\Psi_1} (i=1,2)$, where  $Z_i$ are non-zero complex numbers of non-unit modulus. In addition to this, the metric is invariant under all unitary and anti-unitary transformations due to the Wigner theorem on symmetry transformations.
 
 Next, we consider the infinitesimal distance between two nearby states as given by the Fubini-Study metric for arbitrary time-continuous quantum evolution. All we need is that the quantum state evolves continuously in time. Here, the arbitrary time-continuous evolution means that the dynamics of the system can be unitary, non-unitary, non-Schr{\"o}dinger and even non-linear. For such quantum dynamics, the infinitesimal distance $dS$ between the quantum states $|\Psi(t)\rangle$ and 
$|\Psi(t+dt)\rangle$ on the projective Hilbert space  $\mathcal{P}(\cal{H})$ is given by
 \begin{eqnarray} \label{gf}
 dS^2 =  4 [\langle \dot{\widetilde{\Psi}} (t)|\dot{\widetilde{\Psi}}(t)\rangle-(i\langle \widetilde{\Psi} (t)|\dot{\widetilde{\Psi}}(t)\rangle)^2 ]dt^2,
 \end{eqnarray}
where $|\widetilde{\Psi}(t)\rangle=\frac{|\Psi(t)\rangle}{\norm{|\Psi(t)\rangle}}$ and $\norm{|\Psi(t)\rangle}=\sqrt{\langle\Psi(t)|\Psi(t)\rangle}$.
 
  From the above metric, the evolution speed of the quantum system can be obtained as 
  \begin{eqnarray}\label{QES}
 V(t)=\frac{dS}{dt} = 2 \sqrt{[\langle \dot{\widetilde{\Psi}} (t)|\dot{\widetilde{\Psi}}(t)\rangle-(i\langle \widetilde{\Psi} (t)|\dot{\widetilde{\Psi}}(t)\rangle)^2 ]}.
 \end{eqnarray}
 Integrating the above equation, we obtain the evolution time of the quantum system, which is given by
 \begin{eqnarray}
T=\frac{S}{\overline{V}},
 \end{eqnarray}
 where $\overline{V}=\frac{1}{T} \int^T_0 V(t) dt$. 
 
 Now, the geometry of the projective Hilbert space dictates that the total distance travelled by the state vector as measured by the Fubini-Study metric is always larger than the shortest distance connecting the initial and the final points, i.e., the Fubini-Study distance $S$ is lower bounded by geodesic distance $S_{0}$. Hence, we obtain
 \begin{equation}\label{QSL:Arbi}
     T \geq T_{\rm QSL} = \frac{ S_{0}}{\overline{V}},
 \end{equation}
 where ${S_{0}}= 2 \cos^{-1}({|\langle \widetilde{\Psi}(0)|\widetilde{\Psi}(T)\rangle|})$.

Here, $T_{\rm QSL}$ is the minimum time required for the quantum state to evolve from $|\widetilde{\Psi}(0)\rangle$ to $|\widetilde{\Psi}(T)\rangle$. This is the generalised quantum speed limit (GQSL) for arbitrary time-continuous quantum evolution. The above GQSL bound applies to any quantum time-continuous evolution as long as pure states are mapped to pure states irrespective of unitarity. The evolution can be
unitary, non-unitary, non-linear, completely positive and not completely positive. This is one of the central result of our paper.

Next, we will show that the bound \eqref{QSL:Arbi} reduces to the Mandelstam-Tamm bound \eqref{QSL:unitary} if the dynamics of the quantum system is unitary. If a quantum system evolves by unitary dynamics, then the quantum state $|\Psi(t)\rangle$ obeys the Schr\"odinger  equation
 \begin{eqnarray}
 \label{SWE}
 i \hbar \frac{d}{dt}|\Psi(t)\rangle = H(t)|\Psi(t)\rangle,
 \end{eqnarray}
 where $H(t)$ is the driving Hamiltonian of the quantum system, which is Hermitian.
 
 For the unitary dynamics, the infinitesimal distance $dS$ between two nearby quantum states 
 $|\Psi(t)\rangle$ and 
$|\Psi(t+dt)\rangle$ on the projective Hilbert space  $\mathcal{P}(\cal{H})$ described by Fubini-Study metric~\cite{Anandan1990,akp91,AKP1995} is given by
 \begin{eqnarray}
 \label{FSM}
 dS^2 = 4 [\langle \dot{\Psi} (t)|\dot{\Psi}(t)\rangle-(i\langle \Psi (t)|\dot{\Psi}(t)\rangle)^2 ]dt^2.
 \end{eqnarray}
  Now, using Eq.~\eqref{FSM} and Eq.~\eqref{SWE} we can obtain the quantum evolution speed of the state $|\Psi(t)\rangle$ on the projective Hilbert space $\mathcal{P}(\cal{H})$ as given by
 
 \begin{eqnarray}
V= \frac{dS}{dt} = \frac{ 2 \Delta H(t) }{\hbar},
 \end{eqnarray}
 where $\Delta H(t) = \sqrt{\langle \Psi (t)| H(t)^2|\Psi(t)\rangle 
 - (\langle \Psi (t)| H(t)|\Psi(t)\rangle )^2}$ is the fluctuation in the Hamiltonian of the system. 
 
 Thus, the quantum system moves away from the initial state whose speed is governed by the fluctuation in the Hamiltonian. Integrating the above equation with respect to time gives the total distance travelled by the state as
 \begin{eqnarray}
 S =  \int^T_0 \frac{ 2 \Delta H(t)} {\hbar}dt =\frac{ 2  \overline{\Delta H}(t) T}{\hbar},
 \end{eqnarray}
 where $\overline{\Delta H(t)}=\frac{1}{T} \int^T_0 \Delta H(t) dt$. 
 
 Therefore, the time required for the state to evolve from $|\Psi(0)\rangle$ to
$|\Psi(T)\rangle$ is given by 
  \begin{eqnarray}
 T= \frac{\hbar S}{ 2 \overline{\Delta{H}(t)}}.
 \end{eqnarray}
Using the generalised QSL, we find that 
 \begin{eqnarray}\label{QSL:unitary}
T \geq T_{\rm QSL}= \frac{\hbar S_0}{ 2 \overline{\Delta{H}(t)}},
 \end{eqnarray}
 where ${S_0} = 2 \cos^{-1}(|\langle \Psi (0)|\Psi(T)\rangle|)$.
 
 It is important to note that the above bound was first derived by A.Uhlmann in Ref.~\cite{Uhlmann1992}. If Hamiltonian is time-independent, then we have 
  \begin{eqnarray}
T \geq T_{\rm QSL}= \frac{\hbar S_0}{ 2 {\Delta{H}}}.
 \end{eqnarray}

 This is the famous Mandelstam-Tamm bound~\cite{Mandelstam1945} which suggests that 
$T_{\rm QSL}$ is the minimum time required for the quantum system to evolve from $|\Psi(0)\rangle$ to $|\Psi(T)\rangle$ by unitary evolution.
  
   \subsection{For mixed initial state}
 The quantum speed limit can be generalized for mixed initial states undergoing arbitrary time evolution. Any mixed state of a quantum system can be considered as a reduced state of an enlarged pure entangled state. Consider a quantum system $(S)$ represented by a mixed state $\rho_{S}(t)$ at any time `t' and the ancillary system $(A)$, then the state of the combined system $(S+A)$ is represented by a pure state $|\Psi(t)\rangle_{SA} \in \mathcal{H}_{S}\otimes \mathcal{H}_{A}$ in the enlarged Hilbert space. We can always retrieve the mixed state  $\rho_{S}(t) \in {\mathcal{B}(\mathcal{H}_{S})}$ of $S$ by tracing out $A$. If the density matrix has the spectral decomposition $\rho_{S}(t)=\sum_{i}p_{i}(t)\ket{\phi_{i}
 (t)}\bra{\phi_{i}
 (t)}$, then the purified state is given by

\begin{equation}
    |\Psi(t)\rangle_{SA}=\sum_{i}\sqrt{p_{i}(t)}|\phi_{i}(t)\rangle_{S}|a_{i}\rangle_{A},
\end{equation}
where $p_{i}(t)$ and $\{|\phi_{i}(t)\rangle_{S}\}$ are the eigenvalues and eigenvectors of $\rho_{S}(t)$, respectively and  $\{|a_{i}\rangle_{A}\}$ is time-independent orthonormal basis for the ancillary system. 

Consider the time evolution of the mixed state which generates a curve in the space of density operators, i.e., $\rho(0) \rightarrow \rho(t)$, $0 \le t \le T$. We need not specify what kind of dynamics is obeyed by the density operator. In particular, the density operator may undergo
unitary, completely positive (CP) and non-CP dynamics. For any continuous evolution of the density operator, 
the Fubini-Study metric (\ref{gf}) can be generalized.  For arbitrary time-continuous quantum evolution of the mixed states, we can imagine that the system undergoes continuous time evolution and the ancilla remains stationary. Then the time evolution of the density operator under arbitrary time-continuous dynamics is equivalent to the time evolution of the purified state. In the purified version of the projective Hilbert space, the infinitesimal distance $dS$ between the quantum states $|\Psi(t)\rangle_{SA}$ and 
$|\Psi(t+dt)\rangle_{SA}$ is given by
\begin{eqnarray} 
 dS^2 =  4 [\langle \dot{\widetilde{\Psi}} (t)|\dot{\widetilde{\Psi}}(t)\rangle_{SA}-(i\langle \widetilde{\Psi} (t)|\dot{\widetilde{\Psi}}(t)\rangle_{SA})^2 ]dt^2,
 \end{eqnarray}
where $|\widetilde{\Psi}(t)\rangle_{SA}=\frac{|\Psi(t)\rangle_{SA}}{\norm{|\Psi(t)\rangle_{SA}}}$ and $\norm{|\Psi(t)\rangle_{SA}}=\sqrt{\langle\Psi(t)|\Psi(t)\rangle_{SA}}$.
 
From the above metric, the evolution speed of the quantum system can be obtained as 
 \begin{eqnarray}
 V_{SA}(t)=\frac{dS}{dt} = 2 \sqrt{[\langle \dot{\widetilde{\Psi}} (t)|\dot{\widetilde{\Psi}}(t)\rangle_{SA}-(i\langle \widetilde{\Psi} (t)|\dot{\widetilde{\Psi}}(t)\rangle_{SA})^2 ]}\nonumber\\
 \end{eqnarray}
 
 Integrating the above equation with respect to time, we obtain the evolution time of the quantum system, which is given as
\begin{eqnarray}
T=\frac{S}{\overline{V}_{SA}},
 \end{eqnarray}
where $\overline{V}_{SA}=\frac{1}{T} \int^T_0 {V_{SA}(t)} dt$.
 
 We know that Fubini-Study distance (S) is lower bounded by geodesic distance ${S_{0}}$. Hence, we obtain
 \begin{equation}\label{QSL:Arbi1}
     T \geq T_{\rm QSL} = \frac{ S_{0}}{\bar{V}_{SA}},
 \end{equation}
 where ${S_{0}}=2\cos^{-1}({|\langle\widetilde{\Psi}(0)|\widetilde{\Psi}(T)\rangle_{SA}|})$.
 
 Here, $T_{\rm QSL}$ is the minimum time required for the quantum state to evolve from $|\widetilde{\Psi}(0)\rangle_{SA}$ to $|\widetilde{\Psi}(T)\rangle_{SA}$ by arbitrary time-continuous dynamics. The above GQSL for mixed state is applicable as long as the density matrix evolves continuously in time. The GQSL given in Eq.~\eqref{QSL:Arbi1} reduces to the Eq.~\eqref{QSL:Arbi} for pure state evolution. It is important to note that the above bound~\eqref{QSL:Arbi1} is independent of the choice of purification. A speed limit for an arbitrary time-continuous physical process has also been derived earlier in Ref.~\cite{Taddei2013}. This quantum speed limit applies to most physical processes; however, it requires the estimation of Uhlmann root fidelity (for mixed states) and quantum Fisher information. Therefore, it can be considerably more challenging compared to our bounds. Moreover, from our bounds, one can easily obtain a specific bound for particular dynamics which we mentioned earlier in this paper. Another novelty of our approach is that the Generalized Quantum Speed Limit (GQSL) has been obtained purely from the geometric aspects of quantum evolution. Unlike other derivations of QSL for arbitrary time-continuous dynamics, our result has wide applicability for non-unitary, non-linear, and non-Hermitian quantum evolutions. Even though it may initially appear as a restriction to pure states only, one can eventually apply our QSL to mixed states for open system dynamics by transitioning to a larger Hilbert space, as long as the pure state in the larger Hilbert space evolves continuously in time. Again, within the larger Hilbert space, it need not be restricted to unitary, linear, or any specific quantum evolution.

 \section{GQSL for Non-Hermitian Evolution}\label{QSL:non-Hermition}
We know that, any non-Hermitian Hamiltonian $H$ can always be decomposed into a Hermitian $H_+$ and an anti-Hermitian $H_-$ part as follows 
  \begin{eqnarray}
  \label{h}
H = H_+ + H_- = H_+ - i \Gamma, \label{equation:16}
 \end{eqnarray}
 where
\begin{eqnarray}
\nonumber
H_{\pm} &=& \frac{1}{2}({H}\pm {H}^{\dagger}),
 \end{eqnarray}
 and $\Gamma$ is Hermitian and is commonly known as the decay rate operator.
 
 If $|\Psi(t) \rangle$ (unnormalized) is the time evolved state of the system, then
 \begin{eqnarray}
 \label{ev}
 \nonumber
&& \frac{d }{dt} |\Psi(t) \rangle =- \frac{i}{\hbar} H_+ |\Psi(t) \rangle - \frac{1}{\hbar}
\Gamma |\Psi(t) \rangle, \\ &&\frac{d }{dt}\langle \Psi(t)|  = \frac{i}{\hbar} \langle \Psi(t)|H_+ - \frac{1}{\hbar} \langle \Psi(t)|\Gamma.
 \end{eqnarray}
 
Quantum evolution speed \eqref{QES} for arbitrary Hamiltonian can be written as (see Appendix A)
 \begin{eqnarray}
 V &=&\frac{2}{\hbar}\sqrt{  [
\Delta H_{+}^{2} +\Delta \Gamma^2  +i  \langle [\Gamma,H_+]\rangle
]}.
 \end{eqnarray}
It is interesting to note that the same expression for the speed of evolution for non-Hermitian evolution was obtained in the Refs.~\cite{Barody2012,Impens2021} using non-geometrical approach.
Now using the above evolution speed the bound \eqref{QSL:Arbi} can be written as 
  
 \begin{equation}\label{QSL:non-Her}
     T \geq T_{\rm QSL}= \frac{ \hbar S_{0}}{2\overline{{\sqrt{
\Delta H_{+}^{2} +\Delta \Gamma^2  +i  \langle [\Gamma,H_+]\rangle}}}},
 \end{equation}
where ${S_{0}}=2\cos^{-1}({|\langle \widetilde{\Psi}(0)|\widetilde{\Psi}(T)\rangle|})$.
Note that if $H$ is Hermitian, then $\Gamma=0$ and the above bound reduces to the standard Mandelstam-Tamm bound. It is important to note that the bound~\eqref{QSL:non-Her} can also be derived using uncertainty relation of non-Hermitian operators (without using geometrical approach), for more details see appendix-B. In Ref.~\cite{Bender2007}, it has been argued that evolution time can be made arbitrary small by suitably choosing the parameters of non-Hermitian Hamiltonian in $PT$-symmetric theory. However that is not the case with the lower bound given in Eq.~\eqref{QSL:non-Her}. Moreover, a similar bound is derived in Ref.~\cite{Campo2013} for non-Hermitian dynamics, and they find out their bound not necessarily applicable because of nonlinear nature of $\rho$ in Eq.~\eqref{non-Hermition}. However our bound~\eqref{QSL:non-Her} does not have such limitation.

 \section{QSL For Gain and Loss System}\label{QSL:gain-loss}
The gain (G) and loss (L) system can be thought of as a pair of quantum oscillators G and L that exchange energy coherently at the rate $g$. Furthermore, each oscillator interacts incoherently with its local environment: G pumps energy at a characteristic rate of $\gamma_{G}$ (gain), and  L absorbs energy at a rate of $\gamma_{L}$ (loss). A non-Hermitian Hamiltonian describes the mean-field dynamics of such a quantum system, which is given by~\cite{Roccati2021,Roccati2022}
  \begin{eqnarray}
  \label{gl}
 H  =\left(
\begin{array}{cc}
 -i \gamma_L & g \\
 g & i \gamma_G \\
\end{array}
\right).
\end{eqnarray}

 The eigenvalues of the non-Hermitian Hamiltonian ${H}$ are given by
  \begin{eqnarray}
\lambda_{\pm} = - i \kappa_{-} \pm \delta,
\end{eqnarray}
 where $\delta =  \sqrt{g^2-\kappa^2_{+}}$ and $\kappa_{\pm} = \frac{\gamma_L \pm \gamma_G}{2}$. Here, $\delta$ denotes the transition between the strong and the weak coupling regime. In the strong coupling regime $\delta$ is real as $g > \kappa_{+}$ and in the case of weak coupling  $\delta$ is imaginary as $g < \kappa_{+}$. When $g = \kappa_{+}$, $\delta$ coalesces at the exceptional point.
 
 The non-unitary time evolution operator, which is generated by  $ {H} $ is obtained as

  \begin{eqnarray}
  \nonumber
 \label{u1}
&U_{ {H}}(t) =\\
&\frac{e^{-t \kappa_{-}}}{\delta}
  \nonumber\left(
\begin{array}{cc}
  (\delta  \cos (t \delta )-\kappa_{+} \sin (t \delta )) & - i g \sin (t \delta )\\
 - i g \sin (t \delta ) &  (\delta  \cos (t \delta )+\kappa_{+} \sin (t \delta )) \\
\end{array}
\right).\\
 \end{eqnarray}
 
Let us now assume that the initial state of the quantum system is prepared in a maximally coherent state which is given by 
  \begin{eqnarray}
  \label{ini}
 |\Psi(0)\rangle = \frac{1}{\sqrt{2}} (|0 \rangle  +  |1 \rangle),
  \end{eqnarray}
 and the evolution is governed by the non-unitary operator given in Eq.~\eqref{u1}. The time evolved state of the quantum system is then obtained as
 \begin{eqnarray}
 \nonumber
 \label{gksl}
|\Psi(t)\rangle =  \frac{e^{-t \kappa_{-} }}{\sqrt{2 }\delta N_1  }\left(
\begin{array}{c}
(\delta  \cos (t \delta )-(i g+ \kappa_{+} ) \sin (t \delta )) \\
  (\delta  \cos (t \delta )-(i g - \kappa_{+}) \sin (t \delta )) \\
\end{array}
\right),\\
 \end{eqnarray}
where $N_1 = \frac{e^{ -t \kappa_{-}}}{ \delta } \sqrt{ \left( g^2- \kappa_{+}^2 \cos (2 \delta  t)\right)}$ is the normalization factor.\\

To estimate bound (\ref{ev}), we require the following quantities. The shortest path between the initial state $\ket{\Psi(0)}$ and the final state $\ket{\Psi(T)}$ is given by
  \begin{eqnarray}
 {S_{0}} =2\cos^{-1} \bigg( \bigg|\frac{e^{- T \kappa_{-}} (\delta  \cos (\delta  T)-i g \sin (\delta  T))}{ N \delta } \bigg| \bigg).
 \end{eqnarray}
 In order to find the quantum speed limit of gain and loss system let us decompose the non-Hermitian Hamiltonian $ {H}$ as ${H} = H_+ - i \Gamma$, where
   \begin{eqnarray}
  H_+=\left(
\begin{array}{cc}
 0 & g \\
 g & 0 \\
\end{array}
\right) \   \    {\rm and}  \  \   \Gamma =\left(
\begin{array}{cc}
\gamma_L  & 0 \\
 0 & -\gamma_G \\
\end{array}
\right).
\end{eqnarray}

Now, using the state of the system $|\Psi(t)\rangle$ at an arbitrary time $t$ (Eq.~\eqref{gksl}), the variance of $H_+$ and $\Gamma$, and the expectation value of the commutator of  $\Gamma$ and $H_+$ ($[\Gamma, H_+] = \Gamma H_+ - H_+ \Gamma$) are obtained as follows:
\begin{eqnarray}
\label{h1}
\Delta H_+^2 = g^2 \left(1-\frac{\delta ^4}{\left(g^2-  \kappa_{+}^2 \cos (2 \delta  t)\right)^2}\right),
\end{eqnarray}
\begin{eqnarray}
\label{h11}
\Delta \Gamma^2 =  \kappa_{+}^2-\frac{\delta ^2  \kappa_{+}^4 \sin ^2(2 \delta  t)}{\left(g^2- \kappa_{+}^2 \cos (2 \delta  t)\right)^2},
\end{eqnarray}
and
  \begin{eqnarray}
  \label{h111}
\langle  [\Gamma,H_+]\rangle = \frac{4 i g^2  \kappa_{+}^2 e^{-2 t  \kappa_{-}} \sin ^2(\delta  t)}{\delta ^2},
 \end{eqnarray}
 respectively. Using Eq.~\eqref{h1},  Eq.~\eqref{h11} and  Eq.~\eqref{h111} the evolution speed of gain and loss
system is obtained as
 \begin{eqnarray}
 \nonumber
 V_{GL} &=& \frac{2}{ \left(g^2- \kappa_{+}^2 \cos (2 \delta  t)\right) \sqrt{2}} \bigg[2 g^6-2 g^4  \kappa_{+}^2 \\ &-&g^2 \left(2 \delta ^4+ \kappa_{+}^4\right)-\delta ^2  \kappa_{+}^4+ \kappa_{+}^6 \bigg]^{1/2}.
 \end{eqnarray}
  
 Then using above quantities the quantum speed limit for gain and loss system can be written as
 \begin{equation}\label{gain-loss qsl}
     T \geq \frac{ 2\hbar \cos^{-1} \bigg( \bigg|\frac{e^{-\text{k}_{-} T} (\delta  \cos (\delta  T)-i g \sin (\delta  T))}{ N \delta } \bigg| \bigg)}{ \frac{1}{T} \int^T_0 {  V_{GL}} dt}.
 \end{equation}

\begin{figure}[htp]
    \centering
    \includegraphics[width=8.5cm]{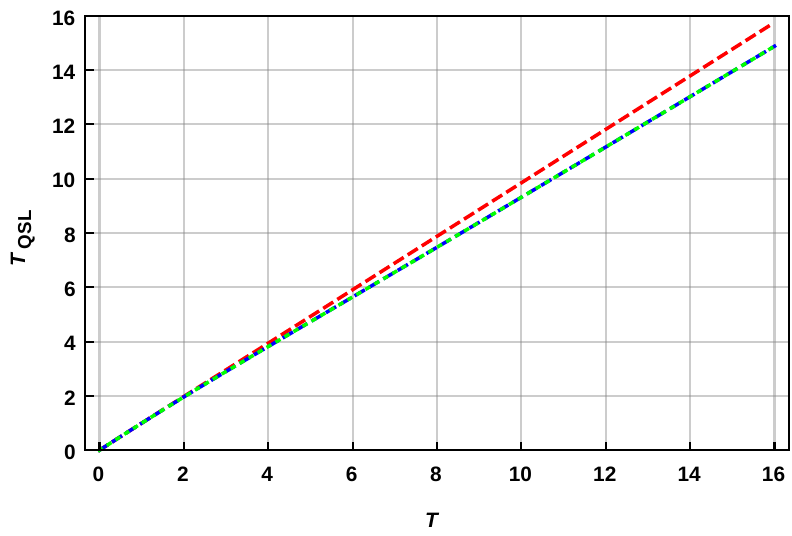}
  \caption{Quantum speed limit $(T_{\rm QSL})$ for gain and loss system is plotted with respect to time $T$ in the weak coupling regime ($g < \kappa_{+}$). The red dashed line is obtained for $g=0.2$, $ \kappa_{+} = 0.6 $ with $\gamma_L = 0.8 $ and $\gamma_G = 0.4 $. The blue dashed line is obtained for the same  $g=0.2$ but  $ \kappa_{+} = 0.3 $ with $\gamma_L = 0.6 $ and $\gamma_G = 0 $. Similarly the green dashed line is obtained for $g=0.2$,  $ \kappa_{+} = 0.3 $ with $\gamma_L = 0 $ and $\gamma_G = 0.6$ }
    \label{fig:GLW}
\end{figure}

\begin{figure}[htp]
    \centering
    \includegraphics[width=8.5cm]{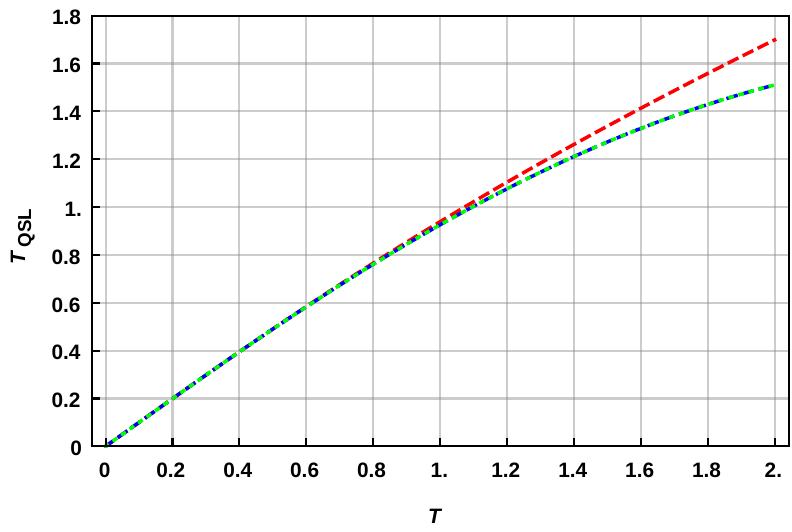}
      \caption{Quantum speed limit ($T_{\rm QSL}$) for gain and loss system is plotted with respect to the evolution time $T$ in the strong coupling regime ($g > \kappa_{+}$). The red dashed line is obtained for $g=0.7 $, $ \kappa_{+} = 0.6 $ with $\gamma_L = 0.8 $ and $\gamma_G = 0.4 $. The blue dashed line is obtained for the same  $g=0.7$ but  $ \kappa_{+} = 0.3 $ with $\gamma_L = 0.6 $ and $\gamma_G = 0 $. Similarly the green dashed line is obtained for $g=0.7$,  $ \kappa_{+} = 0.3 $ with $\gamma_L = 0 $ and $\gamma_G = 0.6 $.}
    \label{fig:GLS}
\end{figure}
\begin{figure}[htp]
    \centering
    \includegraphics[width=8.5cm]{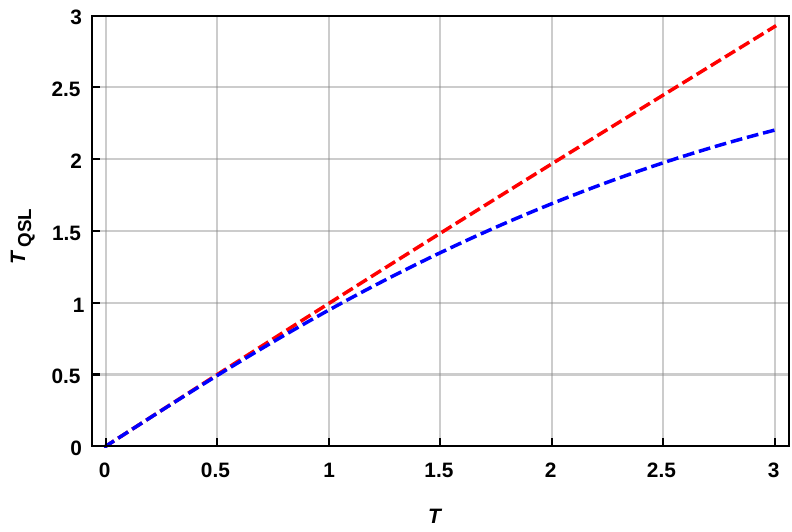}
    \caption{ Quantum speed limit ($T_{\rm QSL}$) for PT-symmetric system is plotted with respect to the evolution time $T$. The red line corresponds to the weak coupling regime at $g=0.2$ and $\kappa_{+} = 0.4  $ with $\gamma =0.4 $ and the blue line corresponds to the strong coupling regime at $g=0.6$  and $\kappa_{+} = 0.4  $ with $\gamma =0.4 $.}
    \label{fig:GLP}
\end{figure}

 In Fig.~\ref{fig:GLW} and  Fig.~\ref{fig:GLS}, we have plotted the quantum speed limit of the gain and loss system in weak and strong coupling regimes, respectively. For gain and loss system, we find that $T_{\rm QSL}$ depends on value of $\kappa_{+}=\frac{\gamma_G+\gamma_L}{2}$. When gain or loss becomes zero it leads to a decrease in the value of $\kappa_{+}$ and as a result $T_{\rm QSL}$ also decreases. In both the cases $\kappa_{+}$ is same, therefore $T_{\rm QSL}$ is identical as we can see in figures ~\ref{fig:GLW} and ~\ref{fig:GLS}. In the weak coupling region, we found that the QSL bound is tight for $\kappa_{+}=0.6$.
 Note that if $\gamma_G = \gamma_ L = \gamma$, ${H}$ in Eq. (\ref{gl}) reduce to $PT$-symmetric Hamiltonian. Substituting $\gamma_G = \gamma_ L = \gamma$ in Eq.~\eqref{gain-loss qsl}, quantum speed limit of $PT$-symmetric quantum system can be obtained. In Fig.~\ref{fig:GLP}, we have plotted the quantum speed limit of the $PT$-symmetric quantum system both for the case of strong and weak coupling regimes. For the $PT$-symmetric case, the QSL bound is tight in the weak coupling region.

\section{QSL For Bethe - Lamb Hamiltonian for two level system}\label{QSL:BH} 
Relativistic corrections and spin-orbit coupling give rise to the splitting of the spectral lines of an atom, formally known as the fine structure \cite{Michelson1887, Sommerfeld1940}. The effects of different external fields on the fine structure of an atom have been extensively studied. Bethe was the first to give a systematic theory of the effect of a static uniform electric field on the fine structure of hydrogen. He considered two degenerate levels $2S_{1/2}$ and $2P_{1/2}$ with different decay constants $\gamma_1$ and $\gamma_2$, respectively, under the influence of a static field $\mathbf{E}_0$. This model can be generalized by allowing for a spacing between the two energy levels, $E_1-E_2=\hbar\omega$. Furthermore, Bethe's original field $\mathbf{E}_0$ when generalised to an ac field, $\mathbf{E}(t)=\mathbf{E}_{0}\sin{\nu t}$ with frequency $\nu$, leads to the Bethe-Lamb Hamiltonian \cite{Lamb1987}.

In this section, we consider a system under this Hamiltonian and apply our formalism to calculate the bound on its evolution time, $T_{\rm QSL}$. The Bethe-Lamb Hamiltonian~\cite{Cui2012} for a general two-level system with radiation damping under the rotating wave approximation is given by
\begin{equation}
    {H}=\left(
\begin{array}{cc}
- \frac{i\gamma_{1}}{2} & \Omega e^{i\Delta t} \\
\Omega e^{-i\Delta t} & - \frac{i\gamma_{2}}{2} \\
\end{array}
\right),
\end{equation}
where all the parameters are taken to be real. Here, $ \gamma_{1}$ and $\gamma_2$ are the decay rates of the excited and ground states respectively, $\Delta=\omega-\nu$ is the laser detuning frequency and $\Omega=\frac{1}{2}\boldsymbol{\mu}_{21}\cdot\mathbf{E}_0$ is the laser Rabi frequency where $\boldsymbol{\mu}_{21}$ is the transition dipole moment for a transition between the initial state and the final state.

We can always decompose the Hamiltonian ${H}$ in the form given in Eq.~(\ref{equation:16}) where 
\begin{align}
H_+  =\left(
\begin{array}{cc}
 0 & \Omega e^{i\Delta t}\\
 \Omega e^{-i\Delta t} & 0 \\
\end{array}
\right)
 \end{align}
 and the decay rate operator is given by 
 \begin{align}
\Gamma  =\left(
\begin{array}{cc}
 \frac{\gamma_1}{2} & 0 \\
 0 & \frac{\gamma_2}{2} \\
\end{array}
\right).
 \end{align}
 This Hamiltonian gives rise to a non-unitary evolution of the system with the corresponding evolution operator $U(t)$, which itself can be decomposed using the Floquet theorem~\cite{Floquet1883,Cui2012} into three parts 
\begin{align}
U(t) = T(t) Z(t) e^{i M t},
\end{align}
where the time-dependent parts are given by
\begin{align}
T(t)=\left(
\begin{array}{cc}
 e^{-\frac{1}{2}\gamma_{1} t} & 0 \\
 0 & e^{-\frac{1}{2}\gamma_{2} t} \\
\end{array}
\right),
\end{align}
\begin{equation}
    Z(t)=\left(
\begin{array}{cc}
 1 & 0 \\
 0 & e^{-it (\Delta -i \gamma )} \\
\end{array}
\right),
\end{equation}
and the time-independent matrix $M$ is given by
\begin{align}
    M=\left(
\begin{array}{cc}
 0 & -\Omega \\
 -\Omega  & (\Delta -i\gamma) \\
\end{array}
\right),
\end{align}
where $\gamma=\frac{1}{2}(\gamma_1-\gamma_2)$.

For the initial system state   $|\Psi(0)\rangle$ \eqref{ini}  at $t=0$, the normalised time evolved state $| \Psi(t) \rangle$ is
\begin{align}
   & |\Psi(t)\rangle=U(t) |\Psi(0)\rangle\nonumber\\ 
    &=\frac{\abs{c_2}}{c_2 \sqrt{(\abs{z_1(t)}^2+\abs{z_2(t)}^2)}}\left( e^{\frac{it\Delta}{2}}z_1(t)|0\rangle +
 e^{-\frac{it\Delta}{2}}z_2(t)|1\rangle
\right),
\end{align}
where 
\begin{eqnarray}
z_1(t)&=&\frac{c_2}{\sqrt{2}}\cosh{\Big(\frac{c_2t}{2}\Big)}-\Big(\frac{1}{\sqrt{2}}(\gamma+i\Delta)+\sqrt{2}i\Omega\Big)\sinh{\Big(\frac{c_2t}{2}\Big)},\nonumber\\
z_2(t)&=&\frac{c_2}{\sqrt{2}}\cosh{\Big(\frac{c_2t}{2}\Big)}+\Big(\frac{1}{\sqrt{2}}(\gamma+i\Delta)-\sqrt{2}i\Omega\Big)\sinh{\Big(\frac{c_2t}{2}\Big)},\nonumber\\
c_2&=&\sqrt{(\gamma +i \Delta )^2-4 \Omega ^2 }.\nonumber
\end{eqnarray}

\begin{figure}[htp]
    \centering
    \includegraphics[width=9cm]{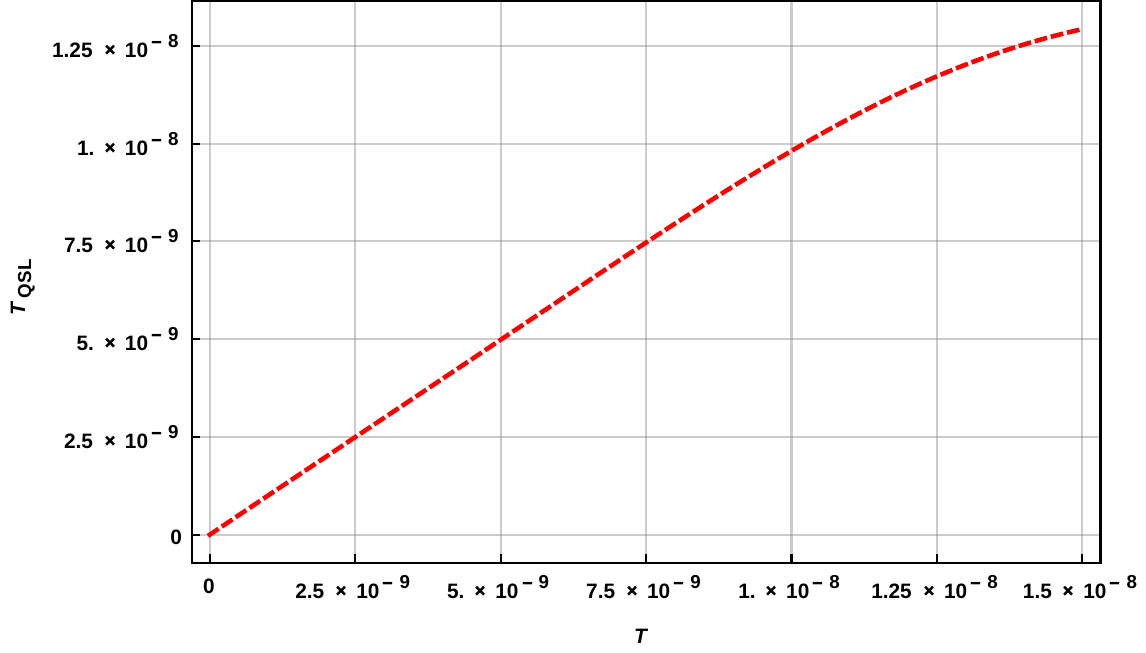}
    \caption{ Quantum speed limit ($T_{\rm QSL}$) for a system under Bethe-Lamb Hamiltonian with the small off-resonance effect as a function of the evolution time $T$. We have taken the parameters from \cite{Lamb1987,Lamb1950,Brunel1998,Han2008}. $\gamma_1^{-1}=10^{-1}$ sec, $\gamma_2^{-1}=1.6*10^{-9}$ sec, $\Delta=180$ MHz and $\Omega=60$ MHz.}
    \label{fig:BLP}
\end{figure}

To estimate the bound~(\ref{QSL:non-Her}) for a system which is described by the Bethe-Lamb Hamiltonian, we need the following quantities:
\begin{eqnarray}
S_0&=& 2 \cos^{-1}{\Bigg(\frac{\abs{c_2}\big(e^{iT\Delta/2}z_1(T)+e^{-iT\Delta/2}z_2(T)\big)}{c_2\sqrt{2\big(\abs{z_1(T)}^2+\abs{z_2(T)}^2\big)}}\Bigg)},\nonumber\\ \\
\Delta H_+^2&=&\Omega^2\Bigg(1-4\frac{\Re{z_{1}(t){z_{2}(t)}^*}^2}{\big(\abs{z_1(t)}^2+\abs{z_2(t)}^2\big)^2}\Bigg),\\ 
\Delta \Gamma^2&=&\frac{\abs{z_1(t)}^2\abs{z_2(t)}^2\gamma^2}{\big(\abs{z_1(t)}^2+\abs{z_2(t)}^2\big)^2},\\ 
\textrm{and} \nonumber\\
i\langle[\Gamma,H_+]\rangle&=&\frac{2\Omega\gamma}{\big(\abs{z_1(t)}^2+\abs{z_2(t)}^2\big)}\Im{z_{1}(t)z_{2}(t)^{*}},
\end{eqnarray}\\
where $\Re\{\}$ and $\Im\{\}$ denote the real and imaginary parts of their arguments, respectively.

We can now evaluate $T_{\rm QSL}$ using Eq. \eqref{QSL:non-Her} where we take $\hbar=1$. In Fig.~\ref{fig:BLP} we plot the numerical estimate of the bound on the evolution time $T_{\rm QSL}$ for the Bethe-Lamb model as a function of the evolution time $T$. We find that the bound is tight for evolution time of the order of the lifetime of the ground state. This is a remarkable prediction of the generalised quantum speed limit
and brings out the power of the geometric approach for arbitrary quantum evolutions beyond the Schr{\"o}dinger evolution.

\section{Conclusions}
 In summary, we have derived the generalised quantum speed limit (GQSL) for systems undergoing arbitrary quantum evolution. Here, arbitrary time-continuous evolution refers to any possible dynamics that the system may undergo, i.e., it can be unitary, non-unitary, completely positive, non-completely positive or even non-linear. The generalised quantum speed limit reduces to the celebrated Mandelstam-Tamm bound when quantum system undergoes unitary evolution. We also obtained a quantum speed limit for non-Hermitian systems. Using our formalism we investigated the quantum speed limit for a non-Hermitian system that generates gain and loss. Our findings show that the minimal evolution time $T_{QSL}$ of the given time independent non-Hermitian system decreases in both the strong and the weak coupling regimes as the average gain and loss parameter decreases. We found that the speed limit is tight in the weak coupling regime. Furthermore, we have found that for the time-dependent Bethe-Lamb Hamiltonian system, the generalised quantum speed limit is actually tight in predicting the life time of the decaying quantum state. In future, we believe that the generalised quantum speed limit (GQSL) will have important applications in quantum computing, quantum metrology, quantum control theory, quantum thermodynamics, charging and discharging of quantum battery, engineering quantum technologies, and a variety of other fields.

We would like to mention that previous studies have derived speed limits for stochastic evolution, to which our speed limits are not applicable~\cite{Pintos2019,Pintos20211}. In Refs.~\cite{Molina2018,Margolus2021,Mackel2022}, speed limits have been derived that do not involve time evolution. Furthermore, there has been a recent extension of the concept of speed limits beyond state evolution, see Refs.~\cite{Pintos2021,Mohan,Mohan21,Carabba2022,Hornedal2023geometricoperator,Pandey22,Pandey2022,Pandey2023BM}.

\section{ACKNOWLEDGMENTS}
D. Thakuria acknowledges the support from Quantum Enabled Science and Technology (QuEST) project from the Department of Science and Technology, India. B. Mohan and A. Srivastav acknowledge the support of the INFOSYS scholarship. A. K. Pati acknowledges the support from the J C Bose grant from the Department of Science and Technology, India.

\appendix
 
\section{Speed of evolution for non-Hermitian quantum system}
 A non-Hermitian Hamiltonian can always be written as the sum of a Hermitian and an anti-Hermitian operator as follows
  \begin{eqnarray}
{H} = H_+ + H_- = H_+ - i \Gamma,
 \end{eqnarray}
 where
$H_+ =  \frac{1}{2}({H}+ {H}^{\dagger})$ and 
$H_- = \frac{1}{2}({H}- {H}^{\dagger}) =-i \Gamma $. 
 Here, both $H_+$ and $\Gamma$ are Hermitian, and $\Gamma$ is commonly known as the decay rate operator.
 
 If $|\Psi(t) \rangle$ (unnormalized) is the time evolved state of the quantum system, then its equation of motion is given by

  \begin{eqnarray}
 \nonumber
&& \frac{d }{dt} |\Psi(t) \rangle =- \frac{i}{\hbar} H_+ |\Psi(t) \rangle - \frac{1}{\hbar}
\Gamma |\Psi(t) \rangle, \\ &&\frac{d }{dt}\langle \Psi(t)|  = \frac{i}{\hbar} \langle \Psi(t)|H_+ - \frac{1}{\hbar} \langle \Psi(t)|\Gamma.
 \end{eqnarray}
 
  The speed of evolution of the quantum system in the projective Hilbert space is given by
  \begin{eqnarray}
 V(t)=\frac{dS}{dt} = 2 \sqrt{[\langle \dot{\widetilde{\Psi}} (t)|\dot{\widetilde{\Psi}}(t)\rangle-(i\langle \widetilde{\Psi}. (t)|\dot{\widetilde{\Psi}}(t)\rangle)^2 ]}.
 \end{eqnarray}
 \begin{widetext}
In order to obtain a simplified form of the evolution speed for non-Hermitian dynamics we require the following quantities:
 \begin{eqnarray}
 \nonumber
 \frac{d}{dt} \bigg( \frac{|\Psi(t)\rangle }{\norm{|\Psi(t)\rangle}} \bigg)&& = \frac{1}{\norm{|\Psi(t)\rangle}^2}\bigg[\norm{|\Psi(t)\rangle}\frac{d |\Psi(t)\rangle}{dt} - |\Psi(t)\rangle \frac{d \norm{|\Psi(t)\rangle}}{dt} \bigg] \\
 && = \frac{1}{ \langle \Psi(t)|\Psi(t)\rangle}\bigg[\sqrt{\langle
\Psi(t)|\Psi(t)\rangle}\frac{d |\Psi(t)\rangle}{dt}+ \frac{\langle \Psi(t)|\Gamma|\Psi(t)\rangle}{\hbar
\sqrt{\langle \Psi(t)|\Psi(t)\rangle}}|\Psi(t)\rangle \bigg],
 \end{eqnarray}

  \begin{eqnarray}
  \label{1t}
 &&\frac{d}{dt} \bigg( \frac{\langle\Psi(t)| }{\norm{\langle\Psi(t)|}} \bigg)\frac{d}{dt} \bigg( \frac{|\Psi(t)\rangle }{\norm{|\Psi(t)\rangle}} \bigg) = \frac{1}{
\langle \Psi(t)|\Psi(t)\rangle}\bigg( \frac{d \langle \Psi(t)|}{dt} \bigg) \bigg( \frac{d |\Psi(t)\rangle}{dt}\bigg)- \frac{1}{\hbar^2} \frac{\langle
\Psi(t)|\Gamma|\Psi(t)\rangle^2}{\langle \Psi(t)|\Psi(t)\rangle^2},
 \end{eqnarray}
 
   \begin{eqnarray}\label{2t}
   \frac{\langle\Psi(t)|}{\norm{\langle\Psi(t)|}}  \bigg | \frac{d}{dt} \bigg(
\frac{|\Psi(t)\rangle}{\norm{|\Psi(t)\rangle}} \bigg)  = \frac{1}{ \langle \Psi(t)|\Psi(t)\rangle}
\bigg[\langle \Psi(t) | \bigg( \frac{d | \Psi(t)\rangle}{dt}\bigg)+ \frac{1}{\hbar}
\langle \Psi(t)|\Gamma|\Psi(t)\rangle\bigg].
 \end{eqnarray}

 Using Eq. (\ref{1t}) and Eq. (\ref{2t}) in Eq. (\ref{gf}) the evolution speed of a system under
non-Hermitian dynamics is obtained as
 \begin{eqnarray}
 \label{nhf}
  \nonumber
 &&V^2 = 2\left(\frac{1}{ \langle \Psi(t)|\Psi(t)\rangle}\bigg( \frac{d \langle \Psi(t)|}{dt}
\bigg) \bigg( \frac{d | \Psi(t)\rangle}{dt}\bigg)+ \frac{1}{ \langle
\Psi(t)|\Psi(t)\rangle^2} \bigg[\bigg(\langle \Psi(t) | \bigg( \frac{d |
\Psi(t)\rangle}{dt}\bigg)\bigg)^2+\frac{2}{\hbar}\langle \Psi(t) | \bigg( \frac{d |
\Psi(t)\rangle}{dt}\bigg) \langle \Psi(t)|\Gamma|\Psi(t)\rangle \bigg]\right),\\
 \end{eqnarray}
 where
 \begin{eqnarray}
 \label{11t}
 \bigg( \frac{d \langle \Psi(t)|}{dt} \bigg) \bigg( \frac{d | \Psi(t)\rangle}{dt}\bigg) =
\frac{1}{\hbar^2}[ \langle \Psi(t)| H_+^2 |\Psi(t)\rangle + \langle
\Psi(t)|\Gamma^2|\Psi(t)\rangle + i  \langle \Psi(t)|[\Gamma H_+ -H_+ \Gamma]]|\Psi(t)\rangle,
 \end{eqnarray}
 and
  \begin{eqnarray}
  \label{21t}
 \bigg[\bigg(\langle \Psi(t) | \bigg( \frac{d |
\Psi(t)\rangle}{dt}\bigg)\bigg)^2+\frac{2}{\hbar}\langle \Psi(t) | \bigg( \frac{d |
\Psi(t)\rangle}{dt}\bigg) \langle \Psi(t)|\Gamma|\Psi(t)\rangle \bigg] = -\frac{1}{\hbar^2}[
\langle \Psi(t)| H_+ |\Psi(t)\rangle^2 + \langle \Psi(t)|\Gamma|\Psi(t)\rangle^2].
 \end{eqnarray}
  Using Eq. (\ref{11t}) and Eq. (\ref{21t}) in Eq. (\ref{nhf}), the evolution speed of the
non-Hermitian quantum system can be re-written as 
 \begin{eqnarray}
 \nonumber
 V = \frac{2}{\hbar}\sqrt{ \bigg[
\Delta H_+^2 +\Delta \Gamma^2  +i  \langle[\Gamma, H_+]\rangle
\bigg]}.
 \end{eqnarray}
 
 \end{widetext}

\section{Derivation of bound~\eqref{QSL:non-Her} using uncertainty relation of non-Hermitian operators}

Let us consider a quantum system which is described by the initial state $\rho(0)=\ketbra{\Psi(0)}$ (normalized), whose time evolution is governed by non-Hermitian driving Hamiltonian $H(t)=H_+-i\Gamma$, where $H_+$ and $\Gamma$ are Hermitian operators. The time evolution of the state of the system at time 't' is given by \cite{Carmichael2007,Barody2012}
\begin{equation}\label{non-Hermition}
  \frac{d\rho(t)}{dt}=\frac{1}{\hbar}(-i[H_+,\rho(t)]-\{\Gamma,\rho(t)\}+2\rho(t)\tr(\rho(t)\Gamma)).
\end{equation}

Then the rate of the change of the expectation value of the time evolved state $\rho(t)$ in the initial state $\rho_0$ is given by
\begin{eqnarray}
  \frac{d\langle\rho(t)\rangle}{dt}&=&\frac{1}{\hbar}(i\langle[H_+,\rho(t)]\rangle-\langle\{\Gamma,\rho(t)\}\rangle+2\langle\rho(t)\rangle\langle\Gamma\rangle)\nonumber\\
  &=&\frac{i}{\hbar}\big(\langle H(t)^\dagger\rho(t)\rangle-\langle\rho(t) H(t)\rangle+\langle\rho(t)\rangle\langle H(t)\rangle\nonumber\\
  &-&\langle\rho(t)\rangle\langle H(t)^\dagger\rangle\big),
\end{eqnarray}
where $\langle\rho(t)\rangle={\bra{\Psi(0)}\rho(t)\ket{\Psi(0)}}$. 

Let us now define $X=\langle H(t)^\dagger\rho(t)\rangle-\langle H(t)^\dagger\rangle\langle\rho(t)\rangle$ for notational convenience. After taking the modulus squared on both sides of the above equation, we obtain
\begin{eqnarray}
    \abs{ \frac{d\langle\rho(t)\rangle}{dt}}^2&=&\frac{2\abs{X}^2-X^2-{X^*}^2\pm2\abs{X}^2}{{\hbar}^2}\nonumber\\
    &=&\frac{4\abs{X}^2-(X+X^*)^2}{{\hbar}^2},
\end{eqnarray}
where $X^*$ is the complex conjugate of $X$. 

Since the second term on the right hand side of the above equation is a positive number it implies that
\begin{equation}
    \abs{ \frac{d\langle\rho(t)\rangle}{dt}}^2\leq\frac{4}{\hbar^2}\abs{\langle H^\dagger\rho(t)\rangle-\langle H^\dagger\rangle \langle\rho(t)\rangle}^2.
\end{equation}
We now use the uncertainty relation for non-Hermitian operators $(\Delta A)^2(\Delta B)^2\geq\abs{\langle A^\dagger B\rangle-\langle A^\dagger\rangle \langle B\rangle}^2$, where $A$ and $B$ are non-Hermitian operators and $(\Delta O)^2=\langle O^\dagger O\rangle-\langle O^\dagger\rangle \langle O \rangle$  \cite{Pati2015,hall2016,mondal2017,yu2019}. In our case $A=H(t)$ and $B=\rho(t)$, then the uncertainty relation becomes $(\Delta H(t))^2(\Delta \rho(t))^2\geq\abs{\langle H^\dagger \rho(t)\rangle-\langle H^\dagger\rangle \langle \rho(t)\rangle}^2$. Now, using this inequality in the above equation, we obtain
\begin{equation}
    \abs{\frac{d\langle\rho(t)\rangle}{dt}}^2\leq\frac{4}{\hbar^2}(\Delta H(t))^2(\Delta \rho(t))^2.
\end{equation}
Taking the square root of the above equation we obtain
\begin{equation}
    \abs{\frac{d\langle\rho(t)\rangle}{dt}}\leq\frac{2}{\hbar}\Delta H(t)\Delta \rho(t).
\end{equation}
Since $\rho(t)$ is a pure state so we have $(\Delta \rho(t))^2=\langle\rho(t)\rangle(1-\langle\rho(t)\rangle)$. The above inequality can be written as 

\begin{equation}
    {\frac{|d\langle\rho(t)\rangle|}{\Delta \rho(t)}}\leq\frac{2}{\hbar}\Delta H(t)dt.
\end{equation}

After integrating the above inequality we obtain 
\begin{eqnarray}
   T\geq  \frac{\hbar S_0}{2\overline{\Delta H(t)}},
\end{eqnarray}
where ${S_{0}}=2\cos^{-1}({|\langle \widetilde{\Psi}(0)|\widetilde{\Psi}(T)\rangle|})$, $\overline{\Delta H(t)}=\frac{1}{T} \int^T_0 \Delta H(t) dt$ and $\Delta H(t)=\sqrt{
\Delta H_{+}^{2} +\Delta \Gamma^2  +i  \langle [\Gamma,H_+]\rangle}$.

\bibliography{qsl}

\begin{thebibliography}{109}%
\makeatletter
\providecommand \@ifxundefined [1]{%
 \@ifx{#1\undefined}
}%
\providecommand \@ifnum [1]{%
 \ifnum #1\expandafter \@firstoftwo
 \else \expandafter \@secondoftwo
 \fi
}%
\providecommand \@ifx [1]{%
 \ifx #1\expandafter \@firstoftwo
 \else \expandafter \@secondoftwo
 \fi
}%
\providecommand \natexlab [1]{#1}%
\providecommand \enquote  [1]{``#1''}%
\providecommand \bibnamefont  [1]{#1}%
\providecommand \bibfnamefont [1]{#1}%
\providecommand \citenamefont [1]{#1}%
\providecommand \href@noop [0]{\@secondoftwo}%
\providecommand \href [0]{\begingroup \@sanitize@url \@href}%
\providecommand \@href[1]{\@@startlink{#1}\@@href}%
\providecommand \@@href[1]{\endgroup#1\@@endlink}%
\providecommand \@sanitize@url [0]{\catcode `\\12\catcode `\$12\catcode
  `\&12\catcode `\#12\catcode `\^12\catcode `\_12\catcode `\%12\relax}%
\providecommand \@@startlink[1]{}%
\providecommand \@@endlink[0]{}%
\providecommand \url  [0]{\begingroup\@sanitize@url \@url }%
\providecommand \@url [1]{\endgroup\@href {#1}{\urlprefix }}%
\providecommand \urlprefix  [0]{URL }%
\providecommand \Eprint [0]{\href }%
\providecommand \doibase [0]{https://doi.org/}%
\providecommand \selectlanguage [0]{\@gobble}%
\providecommand \bibinfo  [0]{\@secondoftwo}%
\providecommand \bibfield  [0]{\@secondoftwo}%
\providecommand \translation [1]{[#1]}%
\providecommand \BibitemOpen [0]{}%
\providecommand \bibitemStop [0]{}%
\providecommand \bibitemNoStop [0]{.\EOS\space}%
\providecommand \EOS [0]{\spacefactor3000\relax}%
\providecommand \BibitemShut  [1]{\csname bibitem#1\endcsname}%
\let\auto@bib@innerbib\@empty
\bibitem [{\citenamefont {Ashhab}\ \emph {et~al.}(2012)\citenamefont {Ashhab},
  \citenamefont {de~Groot},\ and\ \citenamefont {Nori}}]{AGN12}%
  \BibitemOpen
  \bibfield  {author} {\bibinfo {author} {\bibfnamefont {S.}~\bibnamefont
  {Ashhab}}, \bibinfo {author} {\bibfnamefont {P.~C.}\ \bibnamefont
  {de~Groot}},\ and\ \bibinfo {author} {\bibfnamefont {F.}~\bibnamefont
  {Nori}},\ }\bibfield  {title} {\bibinfo {title} {Speed limits for quantum
  gates in multiqubit systems},\ }\href
  {https://doi.org/10.1103/PhysRevA.85.052327} {\bibfield  {journal} {\bibinfo
  {journal} {Physical Review A}\ }\textbf {\bibinfo {volume} {85}},\ \bibinfo
  {pages} {052327} (\bibinfo {year} {2012})}\BibitemShut {NoStop}%
\bibitem [{\citenamefont {Campaioli}\ \emph
  {et~al.}(2018{\natexlab{a}})\citenamefont {Campaioli}, \citenamefont
  {Pollock},\ and\ \citenamefont {Vinjanampathy}}]{F.Campaioli2018}%
  \BibitemOpen
  \bibfield  {author} {\bibinfo {author} {\bibfnamefont {F.}~\bibnamefont
  {Campaioli}}, \bibinfo {author} {\bibfnamefont {F.~A.}\ \bibnamefont
  {Pollock}},\ and\ \bibinfo {author} {\bibfnamefont {S.}~\bibnamefont
  {Vinjanampathy}},\ }\bibinfo {title} {Quantum {B}atteries},\ in\ \href
  {https://doi.org/10.1007/978-3-319-99046-0_8} {\emph {\bibinfo {booktitle}
  {Thermodynamics in the Quantum Regime: Fundamental Aspects and New
  Directions}}},\ \bibinfo {editor} {edited by\ \bibinfo {editor}
  {\bibfnamefont {F.}~\bibnamefont {Binder}}, \bibinfo {editor} {\bibfnamefont
  {L.~A.}\ \bibnamefont {Correa}}, \bibinfo {editor} {\bibfnamefont
  {C.}~\bibnamefont {Gogolin}}, \bibinfo {editor} {\bibfnamefont
  {J.}~\bibnamefont {Anders}},\ and\ \bibinfo {editor} {\bibfnamefont
  {G.}~\bibnamefont {Adesso}}}\ (\bibinfo  {publisher} {Springer International
  Publishing},\ \bibinfo {address} {Cham},\ \bibinfo {year} {2018})\ pp.\
  \bibinfo {pages} {207--225}\BibitemShut {NoStop}%
\bibitem [{\citenamefont {Juli\`a-Farr\'e}\ \emph {et~al.}(2020)\citenamefont
  {Juli\`a-Farr\'e}, \citenamefont {Salamon}, \citenamefont {Riera},
  \citenamefont {Bera},\ and\ \citenamefont {Lewenstein}}]{Manab2020}%
  \BibitemOpen
  \bibfield  {author} {\bibinfo {author} {\bibfnamefont {S.}~\bibnamefont
  {Juli\`a-Farr\'e}}, \bibinfo {author} {\bibfnamefont {T.}~\bibnamefont
  {Salamon}}, \bibinfo {author} {\bibfnamefont {A.}~\bibnamefont {Riera}},
  \bibinfo {author} {\bibfnamefont {M.~N.}\ \bibnamefont {Bera}},\ and\
  \bibinfo {author} {\bibfnamefont {M.}~\bibnamefont {Lewenstein}},\ }\bibfield
   {title} {\bibinfo {title} {Bounds on the capacity and power of quantum
  batteries},\ }\href {https://doi.org/10.1103/PhysRevResearch.2.023113}
  {\bibfield  {journal} {\bibinfo  {journal} {Physical Review Research}\
  }\textbf {\bibinfo {volume} {2}},\ \bibinfo {pages} {023113} (\bibinfo {year}
  {2020})}\BibitemShut {NoStop}%
\bibitem [{\citenamefont {Mohan}\ and\ \citenamefont {Pati}(2021)}]{Mohan2021}%
  \BibitemOpen
  \bibfield  {author} {\bibinfo {author} {\bibfnamefont {B.}~\bibnamefont
  {Mohan}}\ and\ \bibinfo {author} {\bibfnamefont {A.~K.}\ \bibnamefont
  {Pati}},\ }\bibfield  {title} {\bibinfo {title} {Reverse quantum speed limit:
  How slowly a quantum battery can discharge},\ }\href
  {https://doi.org/10.1103/PhysRevA.104.042209} {\bibfield  {journal} {\bibinfo
   {journal} {Physical Review A}\ }\textbf {\bibinfo {volume} {104}},\ \bibinfo
  {pages} {042209} (\bibinfo {year} {2021})}\BibitemShut {NoStop}%
\bibitem [{\citenamefont {Mohan}\ and\ \citenamefont {Pati}(2022)}]{Mohan21}%
  \BibitemOpen
  \bibfield  {author} {\bibinfo {author} {\bibfnamefont {B.}~\bibnamefont
  {Mohan}}\ and\ \bibinfo {author} {\bibfnamefont {A.~K.}\ \bibnamefont
  {Pati}},\ }\bibfield  {title} {\bibinfo {title} {Quantum speed limits for
  observables},\ }\href {https://doi.org/10.1103/PhysRevA.106.042436}
  {\bibfield  {journal} {\bibinfo  {journal} {Physical Review A}\ }\textbf
  {\bibinfo {volume} {106}},\ \bibinfo {pages} {042436} (\bibinfo {year}
  {2022})}\BibitemShut {NoStop}%
\bibitem [{\citenamefont {Caneva}\ \emph {et~al.}(2009)\citenamefont {Caneva},
  \citenamefont {Murphy}, \citenamefont {Calarco}, \citenamefont {Fazio},
  \citenamefont {Montangero}, \citenamefont {Giovannetti},\ and\ \citenamefont
  {Santoro}}]{Caneva2009}%
  \BibitemOpen
  \bibfield  {author} {\bibinfo {author} {\bibfnamefont {T.}~\bibnamefont
  {Caneva}}, \bibinfo {author} {\bibfnamefont {M.}~\bibnamefont {Murphy}},
  \bibinfo {author} {\bibfnamefont {T.}~\bibnamefont {Calarco}}, \bibinfo
  {author} {\bibfnamefont {R.}~\bibnamefont {Fazio}}, \bibinfo {author}
  {\bibfnamefont {S.}~\bibnamefont {Montangero}}, \bibinfo {author}
  {\bibfnamefont {V.}~\bibnamefont {Giovannetti}},\ and\ \bibinfo {author}
  {\bibfnamefont {G.~E.}\ \bibnamefont {Santoro}},\ }\bibfield  {title}
  {\bibinfo {title} {Optimal {C}ontrol at the {Q}uantum {S}peed {L}imit},\
  }\href {https://doi.org/10.1103/PhysRevLett.103.240501} {\bibfield  {journal}
  {\bibinfo  {journal} {Physical review letters}\ }\textbf {\bibinfo {volume}
  {103}},\ \bibinfo {pages} {240501} (\bibinfo {year} {2009})}\BibitemShut
  {NoStop}%
\bibitem [{\citenamefont {Campbell}\ and\ \citenamefont
  {Deffner}(2017)}]{Campbell2017}%
  \BibitemOpen
  \bibfield  {author} {\bibinfo {author} {\bibfnamefont {S.}~\bibnamefont
  {Campbell}}\ and\ \bibinfo {author} {\bibfnamefont {S.}~\bibnamefont
  {Deffner}},\ }\bibfield  {title} {\bibinfo {title} {Trade-{O}ff {B}etween
  {S}peed and {C}ost in {S}hortcuts to {A}diabaticity},\ }\href
  {https://doi.org/10.1103/PhysRevLett.118.100601} {\bibfield  {journal}
  {\bibinfo  {journal} {Physical Review Letters}\ }\textbf {\bibinfo {volume}
  {118}},\ \bibinfo {pages} {100601} (\bibinfo {year} {2017})}\BibitemShut
  {NoStop}%
\bibitem [{\citenamefont {Demkowicz-Dobrza{\'{n}}ski}\ \emph
  {et~al.}(2012)\citenamefont {Demkowicz-Dobrza{\'{n}}ski}, \citenamefont
  {Ko{\l}ody{\'{n}}ski},\ and\ \citenamefont
  {Gu{\c{T}}{\u{a}}}}]{Demkowicz2012}%
  \BibitemOpen
  \bibfield  {author} {\bibinfo {author} {\bibfnamefont {R.}~\bibnamefont
  {Demkowicz-Dobrza{\'{n}}ski}}, \bibinfo {author} {\bibfnamefont
  {J.}~\bibnamefont {Ko{\l}ody{\'{n}}ski}},\ and\ \bibinfo {author}
  {\bibfnamefont {M.}~\bibnamefont {Gu{\c{T}}{\u{a}}}},\ }\bibfield  {title}
  {\bibinfo {title} {The elusive heisenberg limit in quantum-enhanced
  metrology},\ }\href {https://doi.org/10.1038/ncomms2067} {\bibfield
  {journal} {\bibinfo  {journal} {Nature Communications}\ }\textbf {\bibinfo
  {volume} {3}},\ \bibinfo {pages} {1063} (\bibinfo {year} {2012})}\BibitemShut
  {NoStop}%
\bibitem [{\citenamefont {del Campo}\ \emph {et~al.}(2013)\citenamefont {del
  Campo}, \citenamefont {Egusquiza}, \citenamefont {Plenio},\ and\
  \citenamefont {Huelga}}]{Campo2013}%
  \BibitemOpen
  \bibfield  {author} {\bibinfo {author} {\bibfnamefont {A.}~\bibnamefont {del
  Campo}}, \bibinfo {author} {\bibfnamefont {I.~L.}\ \bibnamefont {Egusquiza}},
  \bibinfo {author} {\bibfnamefont {M.~B.}\ \bibnamefont {Plenio}},\ and\
  \bibinfo {author} {\bibfnamefont {S.~F.}\ \bibnamefont {Huelga}},\ }\bibfield
   {title} {\bibinfo {title} {Quantum {S}peed {L}imits in {O}pen {S}ystem
  {D}ynamics},\ }\href {https://doi.org/10.1103/PhysRevLett.110.050403}
  {\bibfield  {journal} {\bibinfo  {journal} {Physical Review Letters}\
  }\textbf {\bibinfo {volume} {110}},\ \bibinfo {pages} {050403} (\bibinfo
  {year} {2013})}\BibitemShut {NoStop}%
\bibitem [{\citenamefont {Tóth}\ and\ \citenamefont
  {Apellaniz}(2014)}]{Toth2014}%
  \BibitemOpen
  \bibfield  {author} {\bibinfo {author} {\bibfnamefont {G.}~\bibnamefont
  {Tóth}}\ and\ \bibinfo {author} {\bibfnamefont {I.}~\bibnamefont
  {Apellaniz}},\ }\bibfield  {title} {\bibinfo {title} {Quantum metrology from
  a quantum information science perspective},\ }\href
  {https://doi.org/10.1088/1751-8113/47/42/424006} {\bibfield  {journal}
  {\bibinfo  {journal} {Journal of Physics A: Mathematical and Theoretical}\
  }\textbf {\bibinfo {volume} {47}},\ \bibinfo {pages} {424006} (\bibinfo
  {year} {2014})}\BibitemShut {NoStop}%
\bibitem [{\citenamefont {Beau}\ and\ \citenamefont {del
  Campo}(2017)}]{Beau2017}%
  \BibitemOpen
  \bibfield  {author} {\bibinfo {author} {\bibfnamefont {M.}~\bibnamefont
  {Beau}}\ and\ \bibinfo {author} {\bibfnamefont {A.}~\bibnamefont {del
  Campo}},\ }\bibfield  {title} {\bibinfo {title} {Nonlinear quantum metrology
  of many-body open systems},\ }\href
  {https://doi.org/10.1103/PhysRevLett.119.010403} {\bibfield  {journal}
  {\bibinfo  {journal} {Physical Review Letters}\ }\textbf {\bibinfo {volume}
  {119}},\ \bibinfo {pages} {010403} (\bibinfo {year} {2017})}\BibitemShut
  {NoStop}%
\bibitem [{\citenamefont {Campbell}\ \emph {et~al.}(2018)\citenamefont
  {Campbell}, \citenamefont {Genoni},\ and\ \citenamefont
  {Deffner}}]{Campbell2018}%
  \BibitemOpen
  \bibfield  {author} {\bibinfo {author} {\bibfnamefont {S.}~\bibnamefont
  {Campbell}}, \bibinfo {author} {\bibfnamefont {M.~G.}\ \bibnamefont
  {Genoni}},\ and\ \bibinfo {author} {\bibfnamefont {S.}~\bibnamefont
  {Deffner}},\ }\bibfield  {title} {\bibinfo {title} {Precision thermometry and
  the quantum speed limit},\ }\href {https://doi.org/10.1088/2058-9565/aaa641}
  {\bibfield  {journal} {\bibinfo  {journal} {Quantum Science and Technology}\
  }\textbf {\bibinfo {volume} {3}},\ \bibinfo {pages} {025002} (\bibinfo {year}
  {2018})}\BibitemShut {NoStop}%
\bibitem [{\citenamefont {Campo}\ \emph {et~al.}(2014)\citenamefont {Campo},
  \citenamefont {Goold},\ and\ \citenamefont {Paternostro}}]{Campo2014}%
  \BibitemOpen
  \bibfield  {author} {\bibinfo {author} {\bibfnamefont {A.~d.}\ \bibnamefont
  {Campo}}, \bibinfo {author} {\bibfnamefont {J.}~\bibnamefont {Goold}},\ and\
  \bibinfo {author} {\bibfnamefont {M.}~\bibnamefont {Paternostro}},\
  }\bibfield  {title} {\bibinfo {title} {More bang for your buck:
  Super-adiabatic quantum engines},\ }\href {https://doi.org/10.1038/srep06208}
  {\bibfield  {journal} {\bibinfo  {journal} {Scientific Reports}\ }\textbf
  {\bibinfo {volume} {4}},\ \bibinfo {pages} {6208} (\bibinfo {year}
  {2014})}\BibitemShut {NoStop}%
\bibitem [{\citenamefont {Mukhopadhyay}\ \emph {et~al.}(2018)\citenamefont
  {Mukhopadhyay}, \citenamefont {Misra}, \citenamefont {Bhattacharya},\ and\
  \citenamefont {Pati}}]{Mukhopadhyay2018}%
  \BibitemOpen
  \bibfield  {author} {\bibinfo {author} {\bibfnamefont {C.}~\bibnamefont
  {Mukhopadhyay}}, \bibinfo {author} {\bibfnamefont {A.}~\bibnamefont {Misra}},
  \bibinfo {author} {\bibfnamefont {S.}~\bibnamefont {Bhattacharya}},\ and\
  \bibinfo {author} {\bibfnamefont {A.~K.}\ \bibnamefont {Pati}},\ }\bibfield
  {title} {\bibinfo {title} {Quantum speed limit constraints on a nanoscale
  autonomous refrigerator},\ }\href
  {https://doi.org/10.1103/PhysRevE.97.062116} {\bibfield  {journal} {\bibinfo
  {journal} {Physical Review E}\ }\textbf {\bibinfo {volume} {97}},\ \bibinfo
  {pages} {062116} (\bibinfo {year} {2018})}\BibitemShut {NoStop}%
\bibitem [{\citenamefont {Mandelstam}\ and\ \citenamefont
  {Tamm}(1945)}]{Mandelstam1945}%
  \BibitemOpen
  \bibfield  {author} {\bibinfo {author} {\bibfnamefont {L.}~\bibnamefont
  {Mandelstam}}\ and\ \bibinfo {author} {\bibfnamefont {I.}~\bibnamefont
  {Tamm}},\ }\bibfield  {title} {\bibinfo {title} {The {U}ncertainty {R}elation
  {B}etween {E}nergy and {T}ime in {N}on-relativistic {Q}uantum {M}echanics},\
  }\href {https://doi.org/10.1007/978-3-642-74626-0_8} {\bibfield  {journal}
  {\bibinfo  {journal} {J. Phys. (USSR)}\ }\textbf {\bibinfo {volume} {9}},\
  \bibinfo {pages} {249} (\bibinfo {year} {1945})}\BibitemShut {NoStop}%
\bibitem [{\citenamefont {Margolus}\ and\ \citenamefont
  {Levitin}(1998)}]{Margolus1998}%
  \BibitemOpen
  \bibfield  {author} {\bibinfo {author} {\bibfnamefont {N.}~\bibnamefont
  {Margolus}}\ and\ \bibinfo {author} {\bibfnamefont {L.~B.}\ \bibnamefont
  {Levitin}},\ }\bibfield  {title} {\bibinfo {title} {The maximum speed of
  dynamical evolution},\ }\href
  {https://doi.org/https://doi.org/10.1016/S0167-2789(98)00054-2} {\bibfield
  {journal} {\bibinfo  {journal} {Physica D: Nonlinear Phenomena}\ }\textbf
  {\bibinfo {volume} {120}},\ \bibinfo {pages} {188} (\bibinfo {year}
  {1998})}\BibitemShut {NoStop}%
\bibitem [{\citenamefont {Anandan}\ and\ \citenamefont
  {Aharonov}(1990)}]{Anandan1990}%
  \BibitemOpen
  \bibfield  {author} {\bibinfo {author} {\bibfnamefont {J.}~\bibnamefont
  {Anandan}}\ and\ \bibinfo {author} {\bibfnamefont {Y.}~\bibnamefont
  {Aharonov}},\ }\bibfield  {title} {\bibinfo {title} {Geometry of quantum
  evolution},\ }\href {https://doi.org/10.1103/PhysRevLett.65.1697} {\bibfield
  {journal} {\bibinfo  {journal} {Physical Review Letters}\ }\textbf {\bibinfo
  {volume} {65}},\ \bibinfo {pages} {1697} (\bibinfo {year}
  {1990})}\BibitemShut {NoStop}%
\bibitem [{\citenamefont {Pati}(1991)}]{akp91}%
  \BibitemOpen
  \bibfield  {author} {\bibinfo {author} {\bibfnamefont {A.~K.}\ \bibnamefont
  {Pati}},\ }\bibfield  {title} {\bibinfo {title} {Relation between
  “phases” and “distance” in quantum evolution},\ }\href
  {https://doi.org/https://doi.org/10.1016/0375-9601(91)90255-7} {\bibfield
  {journal} {\bibinfo  {journal} {Physics Letters A}\ }\textbf {\bibinfo
  {volume} {159}},\ \bibinfo {pages} {105} (\bibinfo {year}
  {1991})}\BibitemShut {NoStop}%
\bibitem [{\citenamefont {Anandan}\ and\ \citenamefont
  {Pati}(1997)}]{Anandan1997}%
  \BibitemOpen
  \bibfield  {author} {\bibinfo {author} {\bibfnamefont {J.~S.}\ \bibnamefont
  {Anandan}}\ and\ \bibinfo {author} {\bibfnamefont {A.~K.}\ \bibnamefont
  {Pati}},\ }\bibfield  {title} {\bibinfo {title} {Geometry of the josephson
  effect},\ }\href
  {https://doi.org/https://doi.org/10.1016/S0375-9601(97)00290-9} {\bibfield
  {journal} {\bibinfo  {journal} {Physics Letters A}\ }\textbf {\bibinfo
  {volume} {231}},\ \bibinfo {pages} {29} (\bibinfo {year} {1997})}\BibitemShut
  {NoStop}%
\bibitem [{\citenamefont {Pati}(1995{\natexlab{a}})}]{Pati1995}%
  \BibitemOpen
  \bibfield  {author} {\bibinfo {author} {\bibfnamefont {A.~K.}\ \bibnamefont
  {Pati}},\ }\bibfield  {title} {\bibinfo {title} {New derivation of the
  geometric phase},\ }\href
  {https://doi.org/https://doi.org/10.1016/0375-9601(95)00299-I} {\bibfield
  {journal} {\bibinfo  {journal} {Physics Letters A}\ }\textbf {\bibinfo
  {volume} {202}},\ \bibinfo {pages} {40} (\bibinfo {year}
  {1995}{\natexlab{a}})}\BibitemShut {NoStop}%
\bibitem [{\citenamefont {Roccati}\ \emph {et~al.}(2021)\citenamefont
  {Roccati}, \citenamefont {Lorenzo}, \citenamefont {Palma}, \citenamefont
  {Landi}, \citenamefont {Brunelli},\ and\ \citenamefont
  {Ciccarello}}]{Roccati2021}%
  \BibitemOpen
  \bibfield  {author} {\bibinfo {author} {\bibfnamefont {F.}~\bibnamefont
  {Roccati}}, \bibinfo {author} {\bibfnamefont {S.}~\bibnamefont {Lorenzo}},
  \bibinfo {author} {\bibfnamefont {G.~M.}\ \bibnamefont {Palma}}, \bibinfo
  {author} {\bibfnamefont {G.~T.}\ \bibnamefont {Landi}}, \bibinfo {author}
  {\bibfnamefont {M.}~\bibnamefont {Brunelli}},\ and\ \bibinfo {author}
  {\bibfnamefont {F.}~\bibnamefont {Ciccarello}},\ }\bibfield  {title}
  {\bibinfo {title} {Quantum correlations in {PT}-symmetric systems},\ }\href
  {https://doi.org/10.1088/2058-9565/abcfcc} {\bibfield  {journal} {\bibinfo
  {journal} {Quantum Science and Technology}\ }\textbf {\bibinfo {volume}
  {6}},\ \bibinfo {pages} {025005} (\bibinfo {year} {2021})}\BibitemShut
  {NoStop}%
\bibitem [{\citenamefont {Roccati}\ \emph {et~al.}(2022)\citenamefont
  {Roccati}, \citenamefont {Palma}, \citenamefont {Bagarello},\ and\
  \citenamefont {Ciccarello}}]{Roccati2022}%
  \BibitemOpen
  \bibfield  {author} {\bibinfo {author} {\bibfnamefont {F.}~\bibnamefont
  {Roccati}}, \bibinfo {author} {\bibfnamefont {G.~M.}\ \bibnamefont {Palma}},
  \bibinfo {author} {\bibfnamefont {F.}~\bibnamefont {Bagarello}},\ and\
  \bibinfo {author} {\bibfnamefont {F.}~\bibnamefont {Ciccarello}},\ }\bibfield
   {title} {\bibinfo {title} {Non-hermitian physics and master equations},\
  }\href {https://doi.org/10.48550/arXiv.2201.05367} {\bibfield  {journal}
  {\bibinfo  {journal} {arXiv:2201.05367}\ } (\bibinfo {year}
  {2022})}\BibitemShut {NoStop}%
\bibitem [{\citenamefont {Mostafazadeh}(2002)}]{Ali2002}%
  \BibitemOpen
  \bibfield  {author} {\bibinfo {author} {\bibfnamefont {A.}~\bibnamefont
  {Mostafazadeh}},\ }\bibfield  {title} {\bibinfo {title} {Pseudo-{H}ermiticity
  versus {P}{T}-symmetry iii: Equivalence of pseudo-{H}ermiticity and the
  presence of antilinear symmetries},\ }\href
  {https://doi.org/10.1063/1.1489072} {\bibfield  {journal} {\bibinfo
  {journal} {Journal of Mathematical Physics}\ }\textbf {\bibinfo {volume}
  {43}},\ \bibinfo {pages} {3944} (\bibinfo {year} {2002})}\BibitemShut
  {NoStop}%
\bibitem [{\citenamefont {Bender}\ \emph {et~al.}(2002)\citenamefont {Bender},
  \citenamefont {Brody},\ and\ \citenamefont {Jones}}]{bender2002}%
  \BibitemOpen
  \bibfield  {author} {\bibinfo {author} {\bibfnamefont {C.~M.}\ \bibnamefont
  {Bender}}, \bibinfo {author} {\bibfnamefont {D.~C.}\ \bibnamefont {Brody}},\
  and\ \bibinfo {author} {\bibfnamefont {H.~F.}\ \bibnamefont {Jones}},\
  }\bibfield  {title} {\bibinfo {title} {Complex {E}xtension of {Q}uantum
  {M}echanics},\ }\href {https://doi.org/10.1103/PhysRevLett.89.270401}
  {\bibfield  {journal} {\bibinfo  {journal} {Physical Review Letters}\
  }\textbf {\bibinfo {volume} {89}},\ \bibinfo {pages} {270401} (\bibinfo
  {year} {2002})}\BibitemShut {NoStop}%
\bibitem [{\citenamefont {Bender}\ \emph {et~al.}(2003)\citenamefont {Bender},
  \citenamefont {Meisinger},\ and\ \citenamefont {Wang}}]{bender2003}%
  \BibitemOpen
  \bibfield  {author} {\bibinfo {author} {\bibfnamefont {C.~M.}\ \bibnamefont
  {Bender}}, \bibinfo {author} {\bibfnamefont {P.~N.}\ \bibnamefont
  {Meisinger}},\ and\ \bibinfo {author} {\bibfnamefont {Q.}~\bibnamefont
  {Wang}},\ }\bibfield  {title} {\bibinfo {title} {Finite-dimensional
  ~~-symmetric {H}amiltonians},\ }\href
  {https://doi.org/10.1088/0305-4470/36/24/314} {\bibfield  {journal} {\bibinfo
   {journal} {Journal of Physics A: Mathematical and General}\ }\textbf
  {\bibinfo {volume} {36}},\ \bibinfo {pages} {6791} (\bibinfo {year}
  {2003})}\BibitemShut {NoStop}%
\bibitem [{\citenamefont {Ding}\ \emph {et~al.}(2021)\citenamefont {Ding},
  \citenamefont {Shi}, \citenamefont {Zhang}, \citenamefont {Shen},
  \citenamefont {Zhang},\ and\ \citenamefont {Zhang}}]{Ding2021}%
  \BibitemOpen
  \bibfield  {author} {\bibinfo {author} {\bibfnamefont {L.}~\bibnamefont
  {Ding}}, \bibinfo {author} {\bibfnamefont {K.}~\bibnamefont {Shi}}, \bibinfo
  {author} {\bibfnamefont {Q.}~\bibnamefont {Zhang}}, \bibinfo {author}
  {\bibfnamefont {D.}~\bibnamefont {Shen}}, \bibinfo {author} {\bibfnamefont
  {X.}~\bibnamefont {Zhang}},\ and\ \bibinfo {author} {\bibfnamefont
  {W.}~\bibnamefont {Zhang}},\ }\bibfield  {title} {\bibinfo {title}
  {Experimental determination of $\mathcal{P}\mathcal{T}$-symmetric exceptional
  points in a single trapped ion},\ }\href
  {https://doi.org/10.1103/PhysRevLett.126.083604} {\bibfield  {journal}
  {\bibinfo  {journal} {Physical Review Letters}\ }\textbf {\bibinfo {volume}
  {126}},\ \bibinfo {pages} {083604} (\bibinfo {year} {2021})}\BibitemShut
  {NoStop}%
\bibitem [{\citenamefont {Wang}\ \emph {et~al.}(2021)\citenamefont {Wang},
  \citenamefont {Zhou}, \citenamefont {Zhang}, \citenamefont {Zhang},
  \citenamefont {Zhang}, \citenamefont {Xie}, \citenamefont {Wu}, \citenamefont
  {Chen}, \citenamefont {Ou}, \citenamefont {Wu}, \citenamefont {Jing},\ and\
  \citenamefont {Chen}}]{Wang2021}%
  \BibitemOpen
  \bibfield  {author} {\bibinfo {author} {\bibfnamefont {W.-C.}\ \bibnamefont
  {Wang}}, \bibinfo {author} {\bibfnamefont {Y.-L.}\ \bibnamefont {Zhou}},
  \bibinfo {author} {\bibfnamefont {H.-L.}\ \bibnamefont {Zhang}}, \bibinfo
  {author} {\bibfnamefont {J.}~\bibnamefont {Zhang}}, \bibinfo {author}
  {\bibfnamefont {M.-C.}\ \bibnamefont {Zhang}}, \bibinfo {author}
  {\bibfnamefont {Y.}~\bibnamefont {Xie}}, \bibinfo {author} {\bibfnamefont
  {C.-W.}\ \bibnamefont {Wu}}, \bibinfo {author} {\bibfnamefont
  {T.}~\bibnamefont {Chen}}, \bibinfo {author} {\bibfnamefont {B.-Q.}\
  \bibnamefont {Ou}}, \bibinfo {author} {\bibfnamefont {W.}~\bibnamefont {Wu}},
  \bibinfo {author} {\bibfnamefont {H.}~\bibnamefont {Jing}},\ and\ \bibinfo
  {author} {\bibfnamefont {P.-X.}\ \bibnamefont {Chen}},\ }\bibfield  {title}
  {\bibinfo {title} {Observation of $\mathcal{PT}$-symmetric quantum coherence
  in a single-ion system},\ }\href
  {https://doi.org/10.1103/PhysRevA.103.L020201} {\bibfield  {journal}
  {\bibinfo  {journal} {Physical Review A}\ }\textbf {\bibinfo {volume}
  {103}},\ \bibinfo {pages} {L020201} (\bibinfo {year} {2021})}\BibitemShut
  {NoStop}%
\bibitem [{\citenamefont {Ding}\ \emph {et~al.}(2022)\citenamefont {Ding},
  \citenamefont {Shi}, \citenamefont {Wang}, \citenamefont {Zhang},
  \citenamefont {Zhu}, \citenamefont {Zhang}, \citenamefont {Yi}, \citenamefont
  {Zhang}, \citenamefont {Zhang},\ and\ \citenamefont {Zhang}}]{Ding2022}%
  \BibitemOpen
  \bibfield  {author} {\bibinfo {author} {\bibfnamefont {L.}~\bibnamefont
  {Ding}}, \bibinfo {author} {\bibfnamefont {K.}~\bibnamefont {Shi}}, \bibinfo
  {author} {\bibfnamefont {Y.}~\bibnamefont {Wang}}, \bibinfo {author}
  {\bibfnamefont {Q.}~\bibnamefont {Zhang}}, \bibinfo {author} {\bibfnamefont
  {C.}~\bibnamefont {Zhu}}, \bibinfo {author} {\bibfnamefont {L.}~\bibnamefont
  {Zhang}}, \bibinfo {author} {\bibfnamefont {J.}~\bibnamefont {Yi}}, \bibinfo
  {author} {\bibfnamefont {S.}~\bibnamefont {Zhang}}, \bibinfo {author}
  {\bibfnamefont {X.}~\bibnamefont {Zhang}},\ and\ \bibinfo {author}
  {\bibfnamefont {W.}~\bibnamefont {Zhang}},\ }\bibfield  {title} {\bibinfo
  {title} {Information retrieval and eigenstate coalescence in a non-hermitian
  quantum system with anti-$\mathcal{PT}$ symmetry},\ }\href
  {https://doi.org/10.1103/PhysRevA.105.L010204} {\bibfield  {journal}
  {\bibinfo  {journal} {Physical Review A}\ }\textbf {\bibinfo {volume}
  {105}},\ \bibinfo {pages} {L010204} (\bibinfo {year} {2022})}\BibitemShut
  {NoStop}%
\bibitem [{\citenamefont {Li}\ \emph {et~al.}(2019)\citenamefont {Li},
  \citenamefont {Harter}, \citenamefont {Liu}, \citenamefont {de~Melo},
  \citenamefont {Joglekar},\ and\ \citenamefont {Luo}}]{Li2019}%
  \BibitemOpen
  \bibfield  {author} {\bibinfo {author} {\bibfnamefont {J.}~\bibnamefont
  {Li}}, \bibinfo {author} {\bibfnamefont {A.~K.}\ \bibnamefont {Harter}},
  \bibinfo {author} {\bibfnamefont {J.}~\bibnamefont {Liu}}, \bibinfo {author}
  {\bibfnamefont {L.}~\bibnamefont {de~Melo}}, \bibinfo {author} {\bibfnamefont
  {Y.~N.}\ \bibnamefont {Joglekar}},\ and\ \bibinfo {author} {\bibfnamefont
  {L.}~\bibnamefont {Luo}},\ }\bibfield  {title} {\bibinfo {title} {Observation
  of parity-time symmetry breaking transitions in a dissipative floquet system
  of ultracold atoms},\ }\href {https://doi.org/10.1038/s41467-019-08596-1}
  {\bibfield  {journal} {\bibinfo  {journal} {Nature Communications}\ }\textbf
  {\bibinfo {volume} {10}},\ \bibinfo {pages} {855} (\bibinfo {year}
  {2019})}\BibitemShut {NoStop}%
\bibitem [{\citenamefont {Jiang}\ \emph {et~al.}(2019)\citenamefont {Jiang},
  \citenamefont {Mei}, \citenamefont {Zuo}, \citenamefont {Zhai}, \citenamefont
  {Li}, \citenamefont {Wen},\ and\ \citenamefont {Du}}]{Jiang2019}%
  \BibitemOpen
  \bibfield  {author} {\bibinfo {author} {\bibfnamefont {Y.}~\bibnamefont
  {Jiang}}, \bibinfo {author} {\bibfnamefont {Y.}~\bibnamefont {Mei}}, \bibinfo
  {author} {\bibfnamefont {Y.}~\bibnamefont {Zuo}}, \bibinfo {author}
  {\bibfnamefont {Y.}~\bibnamefont {Zhai}}, \bibinfo {author} {\bibfnamefont
  {J.}~\bibnamefont {Li}}, \bibinfo {author} {\bibfnamefont {J.}~\bibnamefont
  {Wen}},\ and\ \bibinfo {author} {\bibfnamefont {S.}~\bibnamefont {Du}},\
  }\bibfield  {title} {\bibinfo {title} {Anti-parity-time symmetric optical
  four-wave mixing in cold atoms},\ }\href
  {https://doi.org/10.1103/PhysRevLett.123.193604} {\bibfield  {journal}
  {\bibinfo  {journal} {Physical Review Letters}\ }\textbf {\bibinfo {volume}
  {123}},\ \bibinfo {pages} {193604} (\bibinfo {year} {2019})}\BibitemShut
  {NoStop}%
\bibitem [{\citenamefont {Liu}\ \emph {et~al.}(2021)\citenamefont {Liu},
  \citenamefont {Wu}, \citenamefont {Duan}, \citenamefont {Rong},\ and\
  \citenamefont {Du}}]{Liu2021}%
  \BibitemOpen
  \bibfield  {author} {\bibinfo {author} {\bibfnamefont {W.}~\bibnamefont
  {Liu}}, \bibinfo {author} {\bibfnamefont {Y.}~\bibnamefont {Wu}}, \bibinfo
  {author} {\bibfnamefont {C.-K.}\ \bibnamefont {Duan}}, \bibinfo {author}
  {\bibfnamefont {X.}~\bibnamefont {Rong}},\ and\ \bibinfo {author}
  {\bibfnamefont {J.}~\bibnamefont {Du}},\ }\bibfield  {title} {\bibinfo
  {title} {Dynamically encircling an exceptional point in a real quantum
  system},\ }\href {https://doi.org/10.1103/PhysRevLett.126.170506} {\bibfield
  {journal} {\bibinfo  {journal} {Physical Review Letters}\ }\textbf {\bibinfo
  {volume} {126}},\ \bibinfo {pages} {170506} (\bibinfo {year}
  {2021})}\BibitemShut {NoStop}%
\bibitem [{\citenamefont {Wu}\ \emph {et~al.}(2019)\citenamefont {Wu},
  \citenamefont {Liu}, \citenamefont {Geng}, \citenamefont {Song},
  \citenamefont {Ye}, \citenamefont {Duan}, \citenamefont {Rong},\ and\
  \citenamefont {Du}}]{Wu2019}%
  \BibitemOpen
  \bibfield  {author} {\bibinfo {author} {\bibfnamefont {Y.}~\bibnamefont
  {Wu}}, \bibinfo {author} {\bibfnamefont {W.}~\bibnamefont {Liu}}, \bibinfo
  {author} {\bibfnamefont {J.}~\bibnamefont {Geng}}, \bibinfo {author}
  {\bibfnamefont {X.}~\bibnamefont {Song}}, \bibinfo {author} {\bibfnamefont
  {X.}~\bibnamefont {Ye}}, \bibinfo {author} {\bibfnamefont {C.-K.}\
  \bibnamefont {Duan}}, \bibinfo {author} {\bibfnamefont {X.}~\bibnamefont
  {Rong}},\ and\ \bibinfo {author} {\bibfnamefont {J.}~\bibnamefont {Du}},\
  }\bibfield  {title} {\bibinfo {title} {Observation of parity-time symmetry
  breaking in a single-spin system},\ }\href
  {https://doi.org/https://doi.org/10.1126/science.aaw8205} {\bibfield
  {journal} {\bibinfo  {journal} {Science}\ }\textbf {\bibinfo {volume}
  {364}},\ \bibinfo {pages} {878} (\bibinfo {year} {2019})}\BibitemShut
  {NoStop}%
\bibitem [{\citenamefont {Partanen}\ \emph {et~al.}(2019)\citenamefont
  {Partanen}, \citenamefont {Goetz}, \citenamefont {Tan}, \citenamefont
  {Kohvakka}, \citenamefont {Sevriuk}, \citenamefont {Lake}, \citenamefont
  {Kokkoniemi}, \citenamefont {Ikonen}, \citenamefont {Hazra}, \citenamefont
  {M\"akinen}, \citenamefont {Hyypp\"a}, \citenamefont {Gr\"onberg},
  \citenamefont {Vesterinen}, \citenamefont {Silveri},\ and\ \citenamefont
  {M\"ott\"onen}}]{Part2019}%
  \BibitemOpen
  \bibfield  {author} {\bibinfo {author} {\bibfnamefont {M.}~\bibnamefont
  {Partanen}}, \bibinfo {author} {\bibfnamefont {J.}~\bibnamefont {Goetz}},
  \bibinfo {author} {\bibfnamefont {K.~Y.}\ \bibnamefont {Tan}}, \bibinfo
  {author} {\bibfnamefont {K.}~\bibnamefont {Kohvakka}}, \bibinfo {author}
  {\bibfnamefont {V.}~\bibnamefont {Sevriuk}}, \bibinfo {author} {\bibfnamefont
  {R.~E.}\ \bibnamefont {Lake}}, \bibinfo {author} {\bibfnamefont
  {R.}~\bibnamefont {Kokkoniemi}}, \bibinfo {author} {\bibfnamefont
  {J.}~\bibnamefont {Ikonen}}, \bibinfo {author} {\bibfnamefont
  {D.}~\bibnamefont {Hazra}}, \bibinfo {author} {\bibfnamefont
  {A.}~\bibnamefont {M\"akinen}}, \bibinfo {author} {\bibfnamefont
  {E.}~\bibnamefont {Hyypp\"a}}, \bibinfo {author} {\bibfnamefont
  {L.}~\bibnamefont {Gr\"onberg}}, \bibinfo {author} {\bibfnamefont
  {V.}~\bibnamefont {Vesterinen}}, \bibinfo {author} {\bibfnamefont
  {M.}~\bibnamefont {Silveri}},\ and\ \bibinfo {author} {\bibfnamefont
  {M.}~\bibnamefont {M\"ott\"onen}},\ }\bibfield  {title} {\bibinfo {title}
  {Exceptional points in tunable superconducting resonators},\ }\href
  {https://doi.org/10.1103/PhysRevB.100.134505} {\bibfield  {journal} {\bibinfo
   {journal} {Physical Review B}\ }\textbf {\bibinfo {volume} {100}},\ \bibinfo
  {pages} {134505} (\bibinfo {year} {2019})}\BibitemShut {NoStop}%
\bibitem [{\citenamefont {Levitin}\ and\ \citenamefont
  {Toffoli}(2009)}]{Levitin2009}%
  \BibitemOpen
  \bibfield  {author} {\bibinfo {author} {\bibfnamefont {L.~B.}\ \bibnamefont
  {Levitin}}\ and\ \bibinfo {author} {\bibfnamefont {T.}~\bibnamefont
  {Toffoli}},\ }\bibfield  {title} {\bibinfo {title} {Fundamental {L}imit on
  the {R}ate of {Q}uantum {D}ynamics: The {U}nified {B}ound {I}s {T}ight},\
  }\href {https://doi.org/10.1103/PhysRevLett.103.160502} {\bibfield  {journal}
  {\bibinfo  {journal} {Physical Review Letters}\ }\textbf {\bibinfo {volume}
  {103}},\ \bibinfo {pages} {160502} (\bibinfo {year} {2009})}\BibitemShut
  {NoStop}%
\bibitem [{\citenamefont {Gislason}\ \emph {et~al.}(1985)\citenamefont
  {Gislason}, \citenamefont {Sabelli},\ and\ \citenamefont
  {Wood}}]{Gislason1956}%
  \BibitemOpen
  \bibfield  {author} {\bibinfo {author} {\bibfnamefont {E.~A.}\ \bibnamefont
  {Gislason}}, \bibinfo {author} {\bibfnamefont {N.~H.}\ \bibnamefont
  {Sabelli}},\ and\ \bibinfo {author} {\bibfnamefont {J.~W.}\ \bibnamefont
  {Wood}},\ }\bibfield  {title} {\bibinfo {title} {New form of the time-energy
  uncertainty relation},\ }\href {https://doi.org/10.1103/PhysRevA.31.2078}
  {\bibfield  {journal} {\bibinfo  {journal} {Physical Review A}\ }\textbf
  {\bibinfo {volume} {31}},\ \bibinfo {pages} {2078} (\bibinfo {year}
  {1985})}\BibitemShut {NoStop}%
\bibitem [{\citenamefont {Eberly}\ and\ \citenamefont
  {Singh}(1973)}]{Eberly1973}%
  \BibitemOpen
  \bibfield  {author} {\bibinfo {author} {\bibfnamefont {J.~H.}\ \bibnamefont
  {Eberly}}\ and\ \bibinfo {author} {\bibfnamefont {L.~P.~S.}\ \bibnamefont
  {Singh}},\ }\bibfield  {title} {\bibinfo {title} {Time {O}perators, {P}artial
  {S}tationarity, and the {E}nergy-{T}ime {U}ncertainty {R}elation},\ }\href
  {https://doi.org/10.1103/PhysRevD.7.359} {\bibfield  {journal} {\bibinfo
  {journal} {Physical Review D}\ }\textbf {\bibinfo {volume} {7}},\ \bibinfo
  {pages} {359} (\bibinfo {year} {1973})}\BibitemShut {NoStop}%
\bibitem [{\citenamefont {Bauer}\ and\ \citenamefont
  {Mello}(1978)}]{Bauer1978}%
  \BibitemOpen
  \bibfield  {author} {\bibinfo {author} {\bibfnamefont {M.}~\bibnamefont
  {Bauer}}\ and\ \bibinfo {author} {\bibfnamefont {P.}~\bibnamefont {Mello}},\
  }\bibfield  {title} {\bibinfo {title} {The time-energy uncertainty
  relation},\ }\href
  {https://doi.org/https://doi.org/10.1016/0003-4916(78)90223-3} {\bibfield
  {journal} {\bibinfo  {journal} {Annals of Physics}\ }\textbf {\bibinfo
  {volume} {111}},\ \bibinfo {pages} {38} (\bibinfo {year} {1978})}\BibitemShut
  {NoStop}%
\bibitem [{\citenamefont {Bhattacharyya}(1983)}]{Bhattacharyya1983}%
  \BibitemOpen
  \bibfield  {author} {\bibinfo {author} {\bibfnamefont {K.}~\bibnamefont
  {Bhattacharyya}},\ }\bibfield  {title} {\bibinfo {title} {Quantum decay and
  the {M}andelstam-{T}amm-energy inequality},\ }\href
  {https://doi.org/10.1088/0305-4470/16/13/021} {\bibfield  {journal} {\bibinfo
   {journal} {Journal of Physics A: Mathematical and General}\ }\textbf
  {\bibinfo {volume} {16}},\ \bibinfo {pages} {2993} (\bibinfo {year}
  {1983})}\BibitemShut {NoStop}%
\bibitem [{\citenamefont {Leubner}\ and\ \citenamefont
  {Kiener}(1985)}]{Leubner1985}%
  \BibitemOpen
  \bibfield  {author} {\bibinfo {author} {\bibfnamefont {C.}~\bibnamefont
  {Leubner}}\ and\ \bibinfo {author} {\bibfnamefont {C.}~\bibnamefont
  {Kiener}},\ }\bibfield  {title} {\bibinfo {title} {Improvement of the
  {E}berly-{S}ingh time-energy inequality by combination with the
  {M}andelstam-{T}amm approach},\ }\href
  {https://doi.org/10.1103/PhysRevA.31.483} {\bibfield  {journal} {\bibinfo
  {journal} {Physical Review A}\ }\textbf {\bibinfo {volume} {31}},\ \bibinfo
  {pages} {483} (\bibinfo {year} {1985})}\BibitemShut {NoStop}%
\bibitem [{\citenamefont {Vaidman}(1992)}]{Vaidman1992}%
  \BibitemOpen
  \bibfield  {author} {\bibinfo {author} {\bibfnamefont {L.}~\bibnamefont
  {Vaidman}},\ }\bibfield  {title} {\bibinfo {title} {Minimum time for the
  evolution to an orthogonal quantum state},\ }\href
  {https://doi.org/10.1119/1.16940} {\bibfield  {journal} {\bibinfo  {journal}
  {American journal of physics}\ }\textbf {\bibinfo {volume} {60}},\ \bibinfo
  {pages} {182} (\bibinfo {year} {1992})}\BibitemShut {NoStop}%
\bibitem [{\citenamefont {Uhlmann}(1992)}]{Uhlmann1992}%
  \BibitemOpen
  \bibfield  {author} {\bibinfo {author} {\bibfnamefont {A.}~\bibnamefont
  {Uhlmann}},\ }\bibfield  {title} {\bibinfo {title} {An energy dispersion
  estimate},\ }\href
  {https://doi.org/https://doi.org/10.1016/0375-9601(92)90555-Z} {\bibfield
  {journal} {\bibinfo  {journal} {Physics Letters A}\ }\textbf {\bibinfo
  {volume} {161}},\ \bibinfo {pages} {329} (\bibinfo {year}
  {1992})}\BibitemShut {NoStop}%
\bibitem [{\citenamefont {Uffink}(1993)}]{Uffink1993}%
  \BibitemOpen
  \bibfield  {author} {\bibinfo {author} {\bibfnamefont {J.~B.}\ \bibnamefont
  {Uffink}},\ }\bibfield  {title} {\bibinfo {title} {The rate of evolution of a
  quantum state},\ }\href {https://doi.org/10.1119/1.17368} {\bibfield
  {journal} {\bibinfo  {journal} {American Journal of Physics}\ }\textbf
  {\bibinfo {volume} {61}},\ \bibinfo {pages} {935} (\bibinfo {year}
  {1993})}\BibitemShut {NoStop}%
\bibitem [{\citenamefont {Pfeifer}\ and\ \citenamefont
  {Fr\"ohlich}(1995)}]{Pfeifer1995}%
  \BibitemOpen
  \bibfield  {author} {\bibinfo {author} {\bibfnamefont {P.}~\bibnamefont
  {Pfeifer}}\ and\ \bibinfo {author} {\bibfnamefont {J.}~\bibnamefont
  {Fr\"ohlich}},\ }\bibfield  {title} {\bibinfo {title} {Generalized
  time-energy uncertainty relations and bounds on lifetimes of resonances},\
  }\href {https://doi.org/10.1103/RevModPhys.67.759} {\bibfield  {journal}
  {\bibinfo  {journal} {Reviews of Modern Physics}\ }\textbf {\bibinfo {volume}
  {67}},\ \bibinfo {pages} {759} (\bibinfo {year} {1995})}\BibitemShut
  {NoStop}%
\bibitem [{\citenamefont {Horesh}\ and\ \citenamefont
  {Mann}(1998)}]{Horesh1998}%
  \BibitemOpen
  \bibfield  {author} {\bibinfo {author} {\bibfnamefont {N.}~\bibnamefont
  {Horesh}}\ and\ \bibinfo {author} {\bibfnamefont {A.}~\bibnamefont {Mann}},\
  }\bibfield  {title} {\bibinfo {title} {Intelligent states for the {A}nandan -
  {A}haronov parameter-based uncertainty relation},\ }\href
  {https://doi.org/10.1088/0305-4470/31/36/003} {\bibfield  {journal} {\bibinfo
   {journal} {Journal of Physics A: Mathematical and General}\ }\textbf
  {\bibinfo {volume} {31}},\ \bibinfo {pages} {L609} (\bibinfo {year}
  {1998})}\BibitemShut {NoStop}%
\bibitem [{\citenamefont {Pati}(1999)}]{AKPati1999}%
  \BibitemOpen
  \bibfield  {author} {\bibinfo {author} {\bibfnamefont {A.~K.}\ \bibnamefont
  {Pati}},\ }\bibfield  {title} {\bibinfo {title} {Uncertainty relation of
  {A}nandan–{A}haronov and intelligent states},\ }\href
  {https://doi.org/https://doi.org/10.1016/S0375-9601(99)00701-X} {\bibfield
  {journal} {\bibinfo  {journal} {Physics Letters A}\ }\textbf {\bibinfo
  {volume} {262}},\ \bibinfo {pages} {296} (\bibinfo {year}
  {1999})}\BibitemShut {NoStop}%
\bibitem [{\citenamefont {S\"oderholm}\ \emph {et~al.}(1999)\citenamefont
  {S\"oderholm}, \citenamefont {Bj\"ork}, \citenamefont {Tsegaye},\ and\
  \citenamefont {Trifonov}}]{Soderholm1999}%
  \BibitemOpen
  \bibfield  {author} {\bibinfo {author} {\bibfnamefont {J.}~\bibnamefont
  {S\"oderholm}}, \bibinfo {author} {\bibfnamefont {G.}~\bibnamefont
  {Bj\"ork}}, \bibinfo {author} {\bibfnamefont {T.}~\bibnamefont {Tsegaye}},\
  and\ \bibinfo {author} {\bibfnamefont {A.}~\bibnamefont {Trifonov}},\
  }\bibfield  {title} {\bibinfo {title} {States that minimize the evolution
  time to become an orthogonal state},\ }\href
  {https://doi.org/10.1103/PhysRevA.59.1788} {\bibfield  {journal} {\bibinfo
  {journal} {Physical Review A}\ }\textbf {\bibinfo {volume} {59}},\ \bibinfo
  {pages} {1788} (\bibinfo {year} {1999})}\BibitemShut {NoStop}%
\bibitem [{\citenamefont {Andrecut}\ and\ \citenamefont
  {Ali}(2004)}]{Andrecut2004}%
  \BibitemOpen
  \bibfield  {author} {\bibinfo {author} {\bibfnamefont {M.}~\bibnamefont
  {Andrecut}}\ and\ \bibinfo {author} {\bibfnamefont {M.~K.}\ \bibnamefont
  {Ali}},\ }\bibfield  {title} {\bibinfo {title} {The adiabatic analogue of the
  {M}argolus{\textendash}{L}evitin theorem},\ }\href
  {https://doi.org/10.1088/0305-4470/37/15/l01} {\bibfield  {journal} {\bibinfo
   {journal} {Journal of Physics A: Mathematical and General}\ }\textbf
  {\bibinfo {volume} {37}},\ \bibinfo {pages} {L157} (\bibinfo {year}
  {2004})}\BibitemShut {NoStop}%
\bibitem [{\citenamefont {Gray}\ and\ \citenamefont {Vogt}(2005)}]{Gray2005}%
  \BibitemOpen
  \bibfield  {author} {\bibinfo {author} {\bibfnamefont {J.~E.}\ \bibnamefont
  {Gray}}\ and\ \bibinfo {author} {\bibfnamefont {A.}~\bibnamefont {Vogt}},\
  }\bibfield  {title} {\bibinfo {title} {Mathematical analysis of the
  {M}andelstam--{T}amm time-energy uncertainty principle},\ }\href
  {https://doi.org/10.1063/1.1897164} {\bibfield  {journal} {\bibinfo
  {journal} {Journal of mathematical physics}\ }\textbf {\bibinfo {volume}
  {46}},\ \bibinfo {pages} {052108} (\bibinfo {year} {2005})}\BibitemShut
  {NoStop}%
\bibitem [{\citenamefont {Luo}\ and\ \citenamefont {Zhang}(2005)}]{Luo2005}%
  \BibitemOpen
  \bibfield  {author} {\bibinfo {author} {\bibfnamefont {S.}~\bibnamefont
  {Luo}}\ and\ \bibinfo {author} {\bibfnamefont {Z.}~\bibnamefont {Zhang}},\
  }\bibfield  {title} {\bibinfo {title} {On {D}ecaying {R}ate of {Q}uantum
  {S}tates},\ }\href {https://doi.org/10.1007/s11005-004-5095-4} {\bibfield
  {journal} {\bibinfo  {journal} {Letters in Mathematical Physics}\ }\textbf
  {\bibinfo {volume} {71}},\ \bibinfo {pages} {1} (\bibinfo {year}
  {2005})}\BibitemShut {NoStop}%
\bibitem [{\citenamefont {Zieli{\'n}ski}\ and\ \citenamefont
  {Zych}(2006)}]{Zielinski2006}%
  \BibitemOpen
  \bibfield  {author} {\bibinfo {author} {\bibfnamefont {B.}~\bibnamefont
  {Zieli{\'n}ski}}\ and\ \bibinfo {author} {\bibfnamefont {M.}~\bibnamefont
  {Zych}},\ }\bibfield  {title} {\bibinfo {title} {Generalization of the
  {M}argolus-{L}evitin bound},\ }\href
  {https://doi.org/10.1103/PhysRevA.74.034301} {\bibfield  {journal} {\bibinfo
  {journal} {Physical Review A}\ }\textbf {\bibinfo {volume} {74}},\ \bibinfo
  {pages} {034301} (\bibinfo {year} {2006})}\BibitemShut {NoStop}%
\bibitem [{\citenamefont {Andrews}(2007)}]{Andrews2007}%
  \BibitemOpen
  \bibfield  {author} {\bibinfo {author} {\bibfnamefont {M.}~\bibnamefont
  {Andrews}},\ }\bibfield  {title} {\bibinfo {title} {Bounds to unitary
  evolution},\ }\href {https://doi.org/10.1103/PhysRevA.75.062112} {\bibfield
  {journal} {\bibinfo  {journal} {Physical Review A}\ }\textbf {\bibinfo
  {volume} {75}},\ \bibinfo {pages} {062112} (\bibinfo {year}
  {2007})}\BibitemShut {NoStop}%
\bibitem [{\citenamefont {Yurtsever}(2010)}]{Yurtsever2010}%
  \BibitemOpen
  \bibfield  {author} {\bibinfo {author} {\bibfnamefont {U.}~\bibnamefont
  {Yurtsever}},\ }\bibfield  {title} {\bibinfo {title} {Fundamental limits on
  the speed of evolution of quantum states},\ }\href
  {https://doi.org/10.1088/0031-8949/82/03/035008} {\bibfield  {journal}
  {\bibinfo  {journal} {Physica Scripta}\ }\textbf {\bibinfo {volume} {82}},\
  \bibinfo {pages} {035008} (\bibinfo {year} {2010})}\BibitemShut {NoStop}%
\bibitem [{\citenamefont {Shuang-Shuang}\ \emph {et~al.}(2010)\citenamefont
  {Shuang-Shuang}, \citenamefont {Nan},\ and\ \citenamefont
  {Shun-Long}}]{Fu2010}%
  \BibitemOpen
  \bibfield  {author} {\bibinfo {author} {\bibfnamefont {F.}~\bibnamefont
  {Shuang-Shuang}}, \bibinfo {author} {\bibfnamefont {L.}~\bibnamefont {Nan}},\
  and\ \bibinfo {author} {\bibfnamefont {L.}~\bibnamefont {Shun-Long}},\
  }\bibfield  {title} {\bibinfo {title} {A {N}ote on {F}undamental {L}imit of
  {Q}uantum {D}ynamics {R}ate},\ }\href
  {https://doi.org/10.1088/0253-6102/54/4/15} {\bibfield  {journal} {\bibinfo
  {journal} {Communications in Theoretical Physics}\ }\textbf {\bibinfo
  {volume} {54}},\ \bibinfo {pages} {661} (\bibinfo {year} {2010})}\BibitemShut
  {NoStop}%
\bibitem [{\citenamefont {Zwierz}(2012)}]{Zwierz2012}%
  \BibitemOpen
  \bibfield  {author} {\bibinfo {author} {\bibfnamefont {M.}~\bibnamefont
  {Zwierz}},\ }\bibfield  {title} {\bibinfo {title} {Comment on ``{G}eometric
  derivation of the quantum speed limit''},\ }\href
  {https://doi.org/10.1103/PhysRevA.86.016101} {\bibfield  {journal} {\bibinfo
  {journal} {Physical Review A}\ }\textbf {\bibinfo {volume} {86}},\ \bibinfo
  {pages} {016101} (\bibinfo {year} {2012})}\BibitemShut {NoStop}%
\bibitem [{\citenamefont {Poggi}\ \emph {et~al.}(2013)\citenamefont {Poggi},
  \citenamefont {Lombardo},\ and\ \citenamefont {Wisniacki}}]{Poggi2013}%
  \BibitemOpen
  \bibfield  {author} {\bibinfo {author} {\bibfnamefont {P.~M.}\ \bibnamefont
  {Poggi}}, \bibinfo {author} {\bibfnamefont {F.~C.}\ \bibnamefont
  {Lombardo}},\ and\ \bibinfo {author} {\bibfnamefont {D.~A.}\ \bibnamefont
  {Wisniacki}},\ }\bibfield  {title} {\bibinfo {title} {Quantum speed limit and
  optimal evolution time in a two-level system},\ }\href
  {https://doi.org/10.1209/0295-5075/104/40005} {\bibfield  {journal} {\bibinfo
   {journal} {Europhysics Letters (EPL)}\ }\textbf {\bibinfo {volume} {104}},\
  \bibinfo {pages} {40005} (\bibinfo {year} {2013})}\BibitemShut {NoStop}%
\bibitem [{\citenamefont {Kupferman}\ and\ \citenamefont
  {Reznik}(2008)}]{Kupferman2008}%
  \BibitemOpen
  \bibfield  {author} {\bibinfo {author} {\bibfnamefont {J.}~\bibnamefont
  {Kupferman}}\ and\ \bibinfo {author} {\bibfnamefont {B.}~\bibnamefont
  {Reznik}},\ }\bibfield  {title} {\bibinfo {title} {Entanglement and the speed
  of evolution in mixed states},\ }\href
  {https://doi.org/10.1103/PhysRevA.78.042305} {\bibfield  {journal} {\bibinfo
  {journal} {Physical Review A}\ }\textbf {\bibinfo {volume} {78}},\ \bibinfo
  {pages} {042305} (\bibinfo {year} {2008})}\BibitemShut {NoStop}%
\bibitem [{\citenamefont {Jones}\ and\ \citenamefont {Kok}(2010)}]{Jones2010}%
  \BibitemOpen
  \bibfield  {author} {\bibinfo {author} {\bibfnamefont {P.~J.}\ \bibnamefont
  {Jones}}\ and\ \bibinfo {author} {\bibfnamefont {P.}~\bibnamefont {Kok}},\
  }\bibfield  {title} {\bibinfo {title} {Geometric derivation of the quantum
  speed limit},\ }\href {https://doi.org/10.1103/PhysRevA.82.022107} {\bibfield
   {journal} {\bibinfo  {journal} {Physical Review A}\ }\textbf {\bibinfo
  {volume} {82}},\ \bibinfo {pages} {022107} (\bibinfo {year}
  {2010})}\BibitemShut {NoStop}%
\bibitem [{\citenamefont {Chau}(2010)}]{Chau2010}%
  \BibitemOpen
  \bibfield  {author} {\bibinfo {author} {\bibfnamefont {H.~F.}\ \bibnamefont
  {Chau}},\ }\bibfield  {title} {\bibinfo {title} {Tight upper bound of the
  maximum speed of evolution of a quantum state},\ }\href
  {https://doi.org/10.1103/PhysRevA.81.062133} {\bibfield  {journal} {\bibinfo
  {journal} {Physical Review A}\ }\textbf {\bibinfo {volume} {81}},\ \bibinfo
  {pages} {062133} (\bibinfo {year} {2010})}\BibitemShut {NoStop}%
\bibitem [{\citenamefont {Deffner}\ and\ \citenamefont
  {Lutz}(2013{\natexlab{a}})}]{S.Deffner2013}%
  \BibitemOpen
  \bibfield  {author} {\bibinfo {author} {\bibfnamefont {S.}~\bibnamefont
  {Deffner}}\ and\ \bibinfo {author} {\bibfnamefont {E.}~\bibnamefont {Lutz}},\
  }\bibfield  {title} {\bibinfo {title} {Energy{\textendash}time uncertainty
  relation for driven quantum systems},\ }\href
  {https://doi.org/10.1088/1751-8113/46/33/335302} {\bibfield  {journal}
  {\bibinfo  {journal} {Journal of Physics A: Mathematical and Theoretical}\
  }\textbf {\bibinfo {volume} {46}},\ \bibinfo {pages} {335302} (\bibinfo
  {year} {2013}{\natexlab{a}})}\BibitemShut {NoStop}%
\bibitem [{\citenamefont {Fung}\ and\ \citenamefont {Chau}(2014)}]{Fung2014}%
  \BibitemOpen
  \bibfield  {author} {\bibinfo {author} {\bibfnamefont {C.-H.~F.}\
  \bibnamefont {Fung}}\ and\ \bibinfo {author} {\bibfnamefont {H.}~\bibnamefont
  {Chau}},\ }\bibfield  {title} {\bibinfo {title} {Relation between physical
  time-energy cost of a quantum process and its information fidelity},\ }\href
  {https://doi.org/10.1103/PhysRevA.90.022333} {\bibfield  {journal} {\bibinfo
  {journal} {Physical Review A}\ }\textbf {\bibinfo {volume} {90}},\ \bibinfo
  {pages} {022333} (\bibinfo {year} {2014})}\BibitemShut {NoStop}%
\bibitem [{\citenamefont {Andersson}\ and\ \citenamefont
  {Heydari}(2014)}]{Andersson2014}%
  \BibitemOpen
  \bibfield  {author} {\bibinfo {author} {\bibfnamefont {O.}~\bibnamefont
  {Andersson}}\ and\ \bibinfo {author} {\bibfnamefont {H.}~\bibnamefont
  {Heydari}},\ }\bibfield  {title} {\bibinfo {title} {Quantum speed limits and
  optimal {H}amiltonians for driven systems in mixed states},\ }\href
  {https://doi.org/10.1088/1751-8113/47/21/215301} {\bibfield  {journal}
  {\bibinfo  {journal} {Journal of Physics A: Mathematical and Theoretical}\
  }\textbf {\bibinfo {volume} {47}},\ \bibinfo {pages} {215301} (\bibinfo
  {year} {2014})}\BibitemShut {NoStop}%
\bibitem [{\citenamefont {Mondal}\ \emph {et~al.}(2016)\citenamefont {Mondal},
  \citenamefont {Datta},\ and\ \citenamefont {Sazim}}]{D.Mondal2016}%
  \BibitemOpen
  \bibfield  {author} {\bibinfo {author} {\bibfnamefont {D.}~\bibnamefont
  {Mondal}}, \bibinfo {author} {\bibfnamefont {C.}~\bibnamefont {Datta}},\ and\
  \bibinfo {author} {\bibfnamefont {S.}~\bibnamefont {Sazim}},\ }\bibfield
  {title} {\bibinfo {title} {Quantum coherence sets the quantum speed limit for
  mixed states},\ }\href
  {https://doi.org/https://doi.org/10.1016/j.physleta.2015.12.015} {\bibfield
  {journal} {\bibinfo  {journal} {Physics Letters A}\ }\textbf {\bibinfo
  {volume} {380}},\ \bibinfo {pages} {689} (\bibinfo {year}
  {2016})}\BibitemShut {NoStop}%
\bibitem [{\citenamefont {Mondal}\ and\ \citenamefont
  {Pati}(2016)}]{Mondal2016}%
  \BibitemOpen
  \bibfield  {author} {\bibinfo {author} {\bibfnamefont {D.}~\bibnamefont
  {Mondal}}\ and\ \bibinfo {author} {\bibfnamefont {A.~K.}\ \bibnamefont
  {Pati}},\ }\bibfield  {title} {\bibinfo {title} {Quantum speed limit for
  mixed states using an experimentally realizable metric},\ }\href
  {https://doi.org/https://doi.org/10.1016/j.physleta.2016.02.018} {\bibfield
  {journal} {\bibinfo  {journal} {Physics Letters A}\ }\textbf {\bibinfo
  {volume} {380}},\ \bibinfo {pages} {1395} (\bibinfo {year}
  {2016})}\BibitemShut {NoStop}%
\bibitem [{\citenamefont {Deffner}\ and\ \citenamefont
  {Campbell}(2017)}]{S.Deffner2017}%
  \BibitemOpen
  \bibfield  {author} {\bibinfo {author} {\bibfnamefont {S.}~\bibnamefont
  {Deffner}}\ and\ \bibinfo {author} {\bibfnamefont {S.}~\bibnamefont
  {Campbell}},\ }\bibfield  {title} {\bibinfo {title} {Quantum speed limits:
  from {H}eisenberg's uncertainty principle to optimal quantum control},\
  }\href {https://doi.org/10.1088/1751-8121/aa86c6} {\bibfield  {journal}
  {\bibinfo  {journal} {Journal of Physics A: Mathematical and Theoretical}\
  }\textbf {\bibinfo {volume} {50}},\ \bibinfo {pages} {453001} (\bibinfo
  {year} {2017})}\BibitemShut {NoStop}%
\bibitem [{\citenamefont {Campaioli}\ \emph
  {et~al.}(2018{\natexlab{b}})\citenamefont {Campaioli}, \citenamefont
  {Pollock}, \citenamefont {Binder},\ and\ \citenamefont
  {Modi}}]{Campaioli2018}%
  \BibitemOpen
  \bibfield  {author} {\bibinfo {author} {\bibfnamefont {F.}~\bibnamefont
  {Campaioli}}, \bibinfo {author} {\bibfnamefont {F.~A.}\ \bibnamefont
  {Pollock}}, \bibinfo {author} {\bibfnamefont {F.~C.}\ \bibnamefont
  {Binder}},\ and\ \bibinfo {author} {\bibfnamefont {K.}~\bibnamefont {Modi}},\
  }\bibfield  {title} {\bibinfo {title} {Tightening {Q}uantum {S}peed {L}imits
  for {A}lmost {A}ll {S}tates},\ }\href
  {https://doi.org/10.1103/PhysRevLett.120.060409} {\bibfield  {journal}
  {\bibinfo  {journal} {Physical Review Letters}\ }\textbf {\bibinfo {volume}
  {120}},\ \bibinfo {pages} {060409} (\bibinfo {year}
  {2018}{\natexlab{b}})}\BibitemShut {NoStop}%
\bibitem [{\citenamefont {Giovannetti}\ \emph {et~al.}(2004)\citenamefont
  {Giovannetti}, \citenamefont {Lloyd},\ and\ \citenamefont
  {Maccone}}]{Giovannetti2004}%
  \BibitemOpen
  \bibfield  {author} {\bibinfo {author} {\bibfnamefont {V.}~\bibnamefont
  {Giovannetti}}, \bibinfo {author} {\bibfnamefont {S.}~\bibnamefont {Lloyd}},\
  and\ \bibinfo {author} {\bibfnamefont {L.}~\bibnamefont {Maccone}},\
  }\bibfield  {title} {\bibinfo {title} {The speed limit of quantum unitary
  evolution},\ }\href {https://doi.org/10.1088/1464-4266/6/8/028} {\bibfield
  {journal} {\bibinfo  {journal} {Journal of Optics B: Quantum and
  Semiclassical Optics}\ }\textbf {\bibinfo {volume} {6}},\ \bibinfo {pages}
  {S807} (\bibinfo {year} {2004})}\BibitemShut {NoStop}%
\bibitem [{\citenamefont {Batle}\ \emph {et~al.}(2005)\citenamefont {Batle},
  \citenamefont {Casas}, \citenamefont {Plastino},\ and\ \citenamefont
  {Plastino}}]{Batle2005}%
  \BibitemOpen
  \bibfield  {author} {\bibinfo {author} {\bibfnamefont {J.}~\bibnamefont
  {Batle}}, \bibinfo {author} {\bibfnamefont {M.}~\bibnamefont {Casas}},
  \bibinfo {author} {\bibfnamefont {A.}~\bibnamefont {Plastino}},\ and\
  \bibinfo {author} {\bibfnamefont {A.~R.}\ \bibnamefont {Plastino}},\
  }\bibfield  {title} {\bibinfo {title} {Connection between entanglement and
  the speed of quantum evolution},\ }\href
  {https://doi.org/10.1103/PhysRevA.72.032337} {\bibfield  {journal} {\bibinfo
  {journal} {Physical Review A}\ }\textbf {\bibinfo {volume} {72}},\ \bibinfo
  {pages} {032337} (\bibinfo {year} {2005})}\BibitemShut {NoStop}%
\bibitem [{\citenamefont {Borr\'as}\ \emph {et~al.}(2006)\citenamefont
  {Borr\'as}, \citenamefont {Casas}, \citenamefont {Plastino},\ and\
  \citenamefont {Plastino}}]{Borras2006}%
  \BibitemOpen
  \bibfield  {author} {\bibinfo {author} {\bibfnamefont {A.}~\bibnamefont
  {Borr\'as}}, \bibinfo {author} {\bibfnamefont {M.}~\bibnamefont {Casas}},
  \bibinfo {author} {\bibfnamefont {A.~R.}\ \bibnamefont {Plastino}},\ and\
  \bibinfo {author} {\bibfnamefont {A.}~\bibnamefont {Plastino}},\ }\bibfield
  {title} {\bibinfo {title} {Entanglement and the lower bounds on the speed of
  quantum evolution},\ }\href {https://doi.org/10.1103/PhysRevA.74.022326}
  {\bibfield  {journal} {\bibinfo  {journal} {Physical Review A}\ }\textbf
  {\bibinfo {volume} {74}},\ \bibinfo {pages} {022326} (\bibinfo {year}
  {2006})}\BibitemShut {NoStop}%
\bibitem [{\citenamefont {Zander}\ \emph {et~al.}(2007)\citenamefont {Zander},
  \citenamefont {Plastino}, \citenamefont {Plastino},\ and\ \citenamefont
  {Casas}}]{Zander2007}%
  \BibitemOpen
  \bibfield  {author} {\bibinfo {author} {\bibfnamefont {C.}~\bibnamefont
  {Zander}}, \bibinfo {author} {\bibfnamefont {A.~R.}\ \bibnamefont
  {Plastino}}, \bibinfo {author} {\bibfnamefont {A.}~\bibnamefont {Plastino}},\
  and\ \bibinfo {author} {\bibfnamefont {M.}~\bibnamefont {Casas}},\ }\bibfield
   {title} {\bibinfo {title} {Entanglement and the speed of evolution of
  multi-partite quantum systems},\ }\href
  {https://doi.org/10.1088/1751-8113/40/11/020} {\bibfield  {journal} {\bibinfo
   {journal} {Journal of Physics A: Mathematical and Theoretical}\ }\textbf
  {\bibinfo {volume} {40}},\ \bibinfo {pages} {2861} (\bibinfo {year}
  {2007})}\BibitemShut {NoStop}%
\bibitem [{\citenamefont {Ness}\ \emph {et~al.}(2022)\citenamefont {Ness},
  \citenamefont {Alberti},\ and\ \citenamefont {Sagi}}]{Ness2022}%
  \BibitemOpen
  \bibfield  {author} {\bibinfo {author} {\bibfnamefont {G.}~\bibnamefont
  {Ness}}, \bibinfo {author} {\bibfnamefont {A.}~\bibnamefont {Alberti}},\ and\
  \bibinfo {author} {\bibfnamefont {Y.}~\bibnamefont {Sagi}},\ }\bibfield
  {title} {\bibinfo {title} {Quantum speed limit for states with a bounded
  energy spectrum},\ }\href {https://doi.org/10.1103/PhysRevLett.129.140403}
  {\bibfield  {journal} {\bibinfo  {journal} {Phys. Rev. Lett.}\ }\textbf
  {\bibinfo {volume} {129}},\ \bibinfo {pages} {140403} (\bibinfo {year}
  {2022})}\BibitemShut {NoStop}%
\bibitem [{\citenamefont {Bagchi}\ \emph {et~al.}(2022)\citenamefont {Bagchi},
  \citenamefont {Srivastav},\ and\ \citenamefont {Pati}}]{Bagchi2022}%
  \BibitemOpen
  \bibfield  {author} {\bibinfo {author} {\bibfnamefont {S.}~\bibnamefont
  {Bagchi}}, \bibinfo {author} {\bibfnamefont {A.}~\bibnamefont {Srivastav}},\
  and\ \bibinfo {author} {\bibfnamefont {A.~K.}\ \bibnamefont {Pati}},\
  }\bibfield  {title} {\bibinfo {title} {Quantum speed limit from tighter
  uncertainty relation},\ }\href {https://doi.org/10.48550/arXiv.2211.14561}
  {\bibfield  {journal} {\bibinfo  {journal} {arXiv:2211.14561}\ } (\bibinfo
  {year} {2022})}\BibitemShut {NoStop}%
\bibitem [{\citenamefont {Pati}\ \emph {et~al.}(2023)\citenamefont {Pati},
  \citenamefont {Mohan}, \citenamefont {Sahil},\ and\ \citenamefont
  {Braunstein}}]{Pati2023}%
  \BibitemOpen
  \bibfield  {author} {\bibinfo {author} {\bibfnamefont {A.~K.}\ \bibnamefont
  {Pati}}, \bibinfo {author} {\bibfnamefont {B.}~\bibnamefont {Mohan}},
  \bibinfo {author} {\bibnamefont {Sahil}},\ and\ \bibinfo {author}
  {\bibfnamefont {S.~L.}\ \bibnamefont {Braunstein}},\ }\bibfield  {title}
  {\bibinfo {title} {Exact quantum speed limits},\ }\href
  {https://doi.org/10.48550/arXiv.2305.03839} {\bibfield  {journal} {\bibinfo
  {journal} {arXiv:2305.03839}\ } (\bibinfo {year} {2023})}\BibitemShut
  {NoStop}%
\bibitem [{\citenamefont {Taddei}\ \emph {et~al.}(2013)\citenamefont {Taddei},
  \citenamefont {Escher}, \citenamefont {Davidovich},\ and\ \citenamefont
  {de~Matos~Filho}}]{Taddei2013}%
  \BibitemOpen
  \bibfield  {author} {\bibinfo {author} {\bibfnamefont {M.~M.}\ \bibnamefont
  {Taddei}}, \bibinfo {author} {\bibfnamefont {B.~M.}\ \bibnamefont {Escher}},
  \bibinfo {author} {\bibfnamefont {L.}~\bibnamefont {Davidovich}},\ and\
  \bibinfo {author} {\bibfnamefont {R.~L.}\ \bibnamefont {de~Matos~Filho}},\
  }\bibfield  {title} {\bibinfo {title} {Quantum {S}peed {L}imit for {P}hysical
  {P}rocesses},\ }\href {https://doi.org/10.1103/PhysRevLett.110.050402}
  {\bibfield  {journal} {\bibinfo  {journal} {Physical Review Letters}\
  }\textbf {\bibinfo {volume} {110}},\ \bibinfo {pages} {050402} (\bibinfo
  {year} {2013})}\BibitemShut {NoStop}%
\bibitem [{\citenamefont {Deffner}\ and\ \citenamefont
  {Lutz}(2013{\natexlab{b}})}]{Deffner2013}%
  \BibitemOpen
  \bibfield  {author} {\bibinfo {author} {\bibfnamefont {S.}~\bibnamefont
  {Deffner}}\ and\ \bibinfo {author} {\bibfnamefont {E.}~\bibnamefont {Lutz}},\
  }\bibfield  {title} {\bibinfo {title} {Quantum {S}peed {L}imit for
  {N}on-{M}arkovian {D}ynamics},\ }\href
  {https://doi.org/10.1103/PhysRevLett.111.010402} {\bibfield  {journal}
  {\bibinfo  {journal} {Physical Review Letters}\ }\textbf {\bibinfo {volume}
  {111}},\ \bibinfo {pages} {010402} (\bibinfo {year}
  {2013}{\natexlab{b}})}\BibitemShut {NoStop}%
\bibitem [{\citenamefont {Fung}\ and\ \citenamefont {Chau}(2013)}]{Fung2013}%
  \BibitemOpen
  \bibfield  {author} {\bibinfo {author} {\bibfnamefont {C.-H.~F.}\
  \bibnamefont {Fung}}\ and\ \bibinfo {author} {\bibfnamefont {H.~F.}\
  \bibnamefont {Chau}},\ }\bibfield  {title} {\bibinfo {title} {Time-energy
  measure for quantum processes},\ }\href
  {https://doi.org/10.1103/PhysRevA.88.012307} {\bibfield  {journal} {\bibinfo
  {journal} {Physical Review A}\ }\textbf {\bibinfo {volume} {88}},\ \bibinfo
  {pages} {012307} (\bibinfo {year} {2013})}\BibitemShut {NoStop}%
\bibitem [{\citenamefont {Pires}\ \emph {et~al.}(2016)\citenamefont {Pires},
  \citenamefont {Cianciaruso}, \citenamefont {C\'eleri}, \citenamefont
  {Adesso},\ and\ \citenamefont {Soares-Pinto}}]{Pires2016}%
  \BibitemOpen
  \bibfield  {author} {\bibinfo {author} {\bibfnamefont {D.~P.}\ \bibnamefont
  {Pires}}, \bibinfo {author} {\bibfnamefont {M.}~\bibnamefont {Cianciaruso}},
  \bibinfo {author} {\bibfnamefont {L.~C.}\ \bibnamefont {C\'eleri}}, \bibinfo
  {author} {\bibfnamefont {G.}~\bibnamefont {Adesso}},\ and\ \bibinfo {author}
  {\bibfnamefont {D.~O.}\ \bibnamefont {Soares-Pinto}},\ }\bibfield  {title}
  {\bibinfo {title} {Generalized {G}eometric {Q}uantum {S}peed {L}imits},\
  }\href {https://doi.org/10.1103/PhysRevX.6.021031} {\bibfield  {journal}
  {\bibinfo  {journal} {Physical Review X}\ }\textbf {\bibinfo {volume} {6}},\
  \bibinfo {pages} {021031} (\bibinfo {year} {2016})}\BibitemShut {NoStop}%
\bibitem [{\citenamefont {Deffner}(2020)}]{S.Deffner2020}%
  \BibitemOpen
  \bibfield  {author} {\bibinfo {author} {\bibfnamefont {S.}~\bibnamefont
  {Deffner}},\ }\bibfield  {title} {\bibinfo {title} {Quantum speed limits and
  the maximal rate of information production},\ }\href
  {https://doi.org/10.1103/PhysRevResearch.2.013161} {\bibfield  {journal}
  {\bibinfo  {journal} {Physical Review Research}\ }\textbf {\bibinfo {volume}
  {2}},\ \bibinfo {pages} {013161} (\bibinfo {year} {2020})}\BibitemShut
  {NoStop}%
\bibitem [{\citenamefont {Jing}\ \emph {et~al.}(2016)\citenamefont {Jing},
  \citenamefont {Wu},\ and\ \citenamefont {Del~Campo}}]{Jing2016}%
  \BibitemOpen
  \bibfield  {author} {\bibinfo {author} {\bibfnamefont {J.}~\bibnamefont
  {Jing}}, \bibinfo {author} {\bibfnamefont {L.-A.}\ \bibnamefont {Wu}},\ and\
  \bibinfo {author} {\bibfnamefont {A.}~\bibnamefont {Del~Campo}},\ }\bibfield
  {title} {\bibinfo {title} {Fundamental {S}peed {L}imits to the {G}eneration
  of {Q}uantumness},\ }\href {https://doi.org/10.1038/srep38149} {\bibfield
  {journal} {\bibinfo  {journal} {Scientific Reports}\ }\textbf {\bibinfo
  {volume} {6}},\ \bibinfo {pages} {38149} (\bibinfo {year}
  {2016})}\BibitemShut {NoStop}%
\bibitem [{\citenamefont {Funo}\ \emph {et~al.}(2019)\citenamefont {Funo},
  \citenamefont {Shiraishi},\ and\ \citenamefont {Saito}}]{Funo2019}%
  \BibitemOpen
  \bibfield  {author} {\bibinfo {author} {\bibfnamefont {K.}~\bibnamefont
  {Funo}}, \bibinfo {author} {\bibfnamefont {N.}~\bibnamefont {Shiraishi}},\
  and\ \bibinfo {author} {\bibfnamefont {K.}~\bibnamefont {Saito}},\ }\bibfield
   {title} {\bibinfo {title} {Speed limit for open quantum systems},\ }\href
  {https://doi.org/10.1088/1367-2630/aaf9f5} {\bibfield  {journal} {\bibinfo
  {journal} {New Journal of Physics}\ }\textbf {\bibinfo {volume} {21}},\
  \bibinfo {pages} {013006} (\bibinfo {year} {2019})}\BibitemShut {NoStop}%
\bibitem [{\citenamefont {Ness}\ \emph {et~al.}(2021)\citenamefont {Ness},
  \citenamefont {Lam}, \citenamefont {Alt}, \citenamefont {Meschede},
  \citenamefont {Sagi},\ and\ \citenamefont {Alberti}}]{Ness}%
  \BibitemOpen
  \bibfield  {author} {\bibinfo {author} {\bibfnamefont {G.}~\bibnamefont
  {Ness}}, \bibinfo {author} {\bibfnamefont {M.~R.}\ \bibnamefont {Lam}},
  \bibinfo {author} {\bibfnamefont {W.}~\bibnamefont {Alt}}, \bibinfo {author}
  {\bibfnamefont {D.}~\bibnamefont {Meschede}}, \bibinfo {author}
  {\bibfnamefont {Y.}~\bibnamefont {Sagi}},\ and\ \bibinfo {author}
  {\bibfnamefont {A.}~\bibnamefont {Alberti}},\ }\bibfield  {title} {\bibinfo
  {title} {Observing crossover between quantum speed limits},\ }\href
  {https://doi.org/10.1126/sciadv.abj9119} {\bibfield  {journal} {\bibinfo
  {journal} {Science Advances}\ }\textbf {\bibinfo {volume} {7}},\ \bibinfo
  {pages} {eabj9119} (\bibinfo {year} {2021})}\BibitemShut {NoStop}%
\bibitem [{\citenamefont {Pati}(1995{\natexlab{b}})}]{AKP1995}%
  \BibitemOpen
  \bibfield  {author} {\bibinfo {author} {\bibfnamefont {A.~K.}\ \bibnamefont
  {Pati}},\ }\bibfield  {title} {\bibinfo {title} {Geometric aspects of
  noncyclic quantum evolutions},\ }\href
  {https://doi.org/10.1103/PhysRevA.52.2576} {\bibfield  {journal} {\bibinfo
  {journal} {Physical Review A}\ }\textbf {\bibinfo {volume} {52}},\ \bibinfo
  {pages} {2576} (\bibinfo {year} {1995}{\natexlab{b}})}\BibitemShut {NoStop}%
\bibitem [{\citenamefont {Brody}\ and\ \citenamefont
  {Graefe}(2012)}]{Barody2012}%
  \BibitemOpen
  \bibfield  {author} {\bibinfo {author} {\bibfnamefont {D.~C.}\ \bibnamefont
  {Brody}}\ and\ \bibinfo {author} {\bibfnamefont {E.-M.}\ \bibnamefont
  {Graefe}},\ }\bibfield  {title} {\bibinfo {title} {Mixed-state evolution in
  the presence of gain and loss},\ }\href
  {https://doi.org/10.1103/PhysRevLett.109.230405} {\bibfield  {journal}
  {\bibinfo  {journal} {Physical Review Letters}\ }\textbf {\bibinfo {volume}
  {109}},\ \bibinfo {pages} {230405} (\bibinfo {year} {2012})}\BibitemShut
  {NoStop}%
\bibitem [{\citenamefont {Impens}\ \emph {et~al.}(2021)\citenamefont {Impens},
  \citenamefont {D'Angelis}, \citenamefont {Pinheiro},\ and\ \citenamefont
  {Gu\'ery-Odelin}}]{Impens2021}%
  \BibitemOpen
  \bibfield  {author} {\bibinfo {author} {\bibfnamefont {F.}~\bibnamefont
  {Impens}}, \bibinfo {author} {\bibfnamefont {F.~M.}\ \bibnamefont
  {D'Angelis}}, \bibinfo {author} {\bibfnamefont {F.~A.}\ \bibnamefont
  {Pinheiro}},\ and\ \bibinfo {author} {\bibfnamefont {D.}~\bibnamefont
  {Gu\'ery-Odelin}},\ }\bibfield  {title} {\bibinfo {title} {Time scaling and
  quantum speed limit in non-hermitian hamiltonians},\ }\href
  {https://doi.org/10.1103/PhysRevA.104.052620} {\bibfield  {journal} {\bibinfo
   {journal} {Physical Review A}\ }\textbf {\bibinfo {volume} {104}},\ \bibinfo
  {pages} {052620} (\bibinfo {year} {2021})}\BibitemShut {NoStop}%
\bibitem [{\citenamefont {Bender}\ \emph {et~al.}(2007)\citenamefont {Bender},
  \citenamefont {Brody}, \citenamefont {Jones},\ and\ \citenamefont
  {Meister}}]{Bender2007}%
  \BibitemOpen
  \bibfield  {author} {\bibinfo {author} {\bibfnamefont {C.~M.}\ \bibnamefont
  {Bender}}, \bibinfo {author} {\bibfnamefont {D.~C.}\ \bibnamefont {Brody}},
  \bibinfo {author} {\bibfnamefont {H.~F.}\ \bibnamefont {Jones}},\ and\
  \bibinfo {author} {\bibfnamefont {B.~K.}\ \bibnamefont {Meister}},\
  }\bibfield  {title} {\bibinfo {title} {Faster than hermitian quantum
  mechanics},\ }\href {https://doi.org/10.1103/PhysRevLett.98.040403}
  {\bibfield  {journal} {\bibinfo  {journal} {Physical Review Letters}\
  }\textbf {\bibinfo {volume} {98}},\ \bibinfo {pages} {040403} (\bibinfo
  {year} {2007})}\BibitemShut {NoStop}%
\bibitem [{\citenamefont {Michelson}\ and\ \citenamefont
  {Morley}(1887)}]{Michelson1887}%
  \BibitemOpen
  \bibfield  {author} {\bibinfo {author} {\bibfnamefont {A.~A.}\ \bibnamefont
  {Michelson}}\ and\ \bibinfo {author} {\bibfnamefont {E.~W.}\ \bibnamefont
  {Morley}},\ }\bibfield  {title} {\bibinfo {title} {On a method of making the
  wave-length of sodium light the actual and practical standard of length},\
  }\href
  {https://babel.hathitrust.org/cgi/pt?id=coo.31924084352636&view=1up&seq=461&skin=2021}
  {\bibfield  {journal} {\bibinfo  {journal} {American Journal of Science
  (1880-1910)}\ }\textbf {\bibinfo {volume} {34}},\ \bibinfo {pages} {427}
  (\bibinfo {year} {1887})}\BibitemShut {NoStop}%
\bibitem [{\citenamefont {Sommerfeld}(1940)}]{Sommerfeld1940}%
  \BibitemOpen
  \bibfield  {author} {\bibinfo {author} {\bibfnamefont {A.}~\bibnamefont
  {Sommerfeld}},\ }\bibfield  {title} {\bibinfo {title} {On the fine structure
  of the hydrogen lines history and current state of theory},\ }\href
  {https://doi.org/https://doi.org/10.1007/BF01490583} {\bibfield  {journal}
  {\bibinfo  {journal} {science}\ }\textbf {\bibinfo {volume} {28}},\ \bibinfo
  {pages} {417} (\bibinfo {year} {1940})}\BibitemShut {NoStop}%
\bibitem [{\citenamefont {Lamb}\ \emph {et~al.}(1987)\citenamefont {Lamb},
  \citenamefont {Schlicher},\ and\ \citenamefont {Scully}}]{Lamb1987}%
  \BibitemOpen
  \bibfield  {author} {\bibinfo {author} {\bibfnamefont {W.~E.}\ \bibnamefont
  {Lamb}}, \bibinfo {author} {\bibfnamefont {R.~R.}\ \bibnamefont
  {Schlicher}},\ and\ \bibinfo {author} {\bibfnamefont {M.~O.}\ \bibnamefont
  {Scully}},\ }\bibfield  {title} {\bibinfo {title} {Matter-field interaction
  in atomic physics and quantum optics},\ }\href
  {https://doi.org/10.1103/PhysRevA.36.2763} {\bibfield  {journal} {\bibinfo
  {journal} {Physical Review A}\ }\textbf {\bibinfo {volume} {36}},\ \bibinfo
  {pages} {2763} (\bibinfo {year} {1987})}\BibitemShut {NoStop}%
\bibitem [{\citenamefont {Cui}\ and\ \citenamefont {Zheng}(2012)}]{Cui2012}%
  \BibitemOpen
  \bibfield  {author} {\bibinfo {author} {\bibfnamefont {X.-D.}\ \bibnamefont
  {Cui}}\ and\ \bibinfo {author} {\bibfnamefont {Y.}~\bibnamefont {Zheng}},\
  }\bibfield  {title} {\bibinfo {title} {Geometric phases in non-hermitian
  quantum mechanics},\ }\href {https://doi.org/10.1103/PhysRevA.86.064104}
  {\bibfield  {journal} {\bibinfo  {journal} {Physical Review A}\ }\textbf
  {\bibinfo {volume} {86}},\ \bibinfo {pages} {064104} (\bibinfo {year}
  {2012})}\BibitemShut {NoStop}%
\bibitem [{\citenamefont {Floquet}(1883)}]{Floquet1883}%
  \BibitemOpen
  \bibfield  {author} {\bibinfo {author} {\bibfnamefont {G.}~\bibnamefont
  {Floquet}},\ }\bibfield  {title} {\bibinfo {title} {On the linear
  {\`e}differential equations {\`with} periodic coefficients},\ }in\ \href@noop
  {} {\emph {\bibinfo {booktitle} {Scientific Annals of the {\'School}Normal
  Superior}}},\ Vol.~\bibinfo {volume} {12}\ (\bibinfo {year} {1883})\ pp.\
  \bibinfo {pages} {47--88}\BibitemShut {NoStop}%
\bibitem [{\citenamefont {Lamb}\ and\ \citenamefont
  {Retherford}(1950)}]{Lamb1950}%
  \BibitemOpen
  \bibfield  {author} {\bibinfo {author} {\bibfnamefont {W.~E.}\ \bibnamefont
  {Lamb}}\ and\ \bibinfo {author} {\bibfnamefont {R.~C.}\ \bibnamefont
  {Retherford}},\ }\bibfield  {title} {\bibinfo {title} {Fine structure of the
  hydrogen atom. part i},\ }\href {https://doi.org/10.1103/PhysRev.79.549}
  {\bibfield  {journal} {\bibinfo  {journal} {Physical Review}\ }\textbf
  {\bibinfo {volume} {79}},\ \bibinfo {pages} {549} (\bibinfo {year}
  {1950})}\BibitemShut {NoStop}%
\bibitem [{\citenamefont {Brunel}\ \emph {et~al.}(1998)\citenamefont {Brunel},
  \citenamefont {Lounis}, \citenamefont {Tamarat},\ and\ \citenamefont
  {Orrit}}]{Brunel1998}%
  \BibitemOpen
  \bibfield  {author} {\bibinfo {author} {\bibfnamefont {C.}~\bibnamefont
  {Brunel}}, \bibinfo {author} {\bibfnamefont {B.}~\bibnamefont {Lounis}},
  \bibinfo {author} {\bibfnamefont {P.}~\bibnamefont {Tamarat}},\ and\ \bibinfo
  {author} {\bibfnamefont {M.}~\bibnamefont {Orrit}},\ }\bibfield  {title}
  {\bibinfo {title} {Rabi resonances of a single molecule driven by rf and
  laser fields},\ }\href {https://doi.org/10.1103/PhysRevLett.81.2679}
  {\bibfield  {journal} {\bibinfo  {journal} {Physical Review Letters}\
  }\textbf {\bibinfo {volume} {81}},\ \bibinfo {pages} {2679} (\bibinfo {year}
  {1998})}\BibitemShut {NoStop}%
\bibitem [{\citenamefont {Han}\ and\ \citenamefont {Zheng}(2008)}]{Han2008}%
  \BibitemOpen
  \bibfield  {author} {\bibinfo {author} {\bibfnamefont {B.}~\bibnamefont
  {Han}}\ and\ \bibinfo {author} {\bibfnamefont {Y.}~\bibnamefont {Zheng}},\
  }\bibfield  {title} {\bibinfo {title} {Control of the photon emission of
  single molecules in a solid matrix},\ }\href
  {https://doi.org/10.1103/PhysRevA.78.015402} {\bibfield  {journal} {\bibinfo
  {journal} {Physical Review A}\ }\textbf {\bibinfo {volume} {78}},\ \bibinfo
  {pages} {015402} (\bibinfo {year} {2008})}\BibitemShut {NoStop}%
\bibitem [{\citenamefont {García-Pintos}\ and\ \citenamefont {del
  Campo}(2019)}]{Pintos2019}%
  \BibitemOpen
  \bibfield  {author} {\bibinfo {author} {\bibfnamefont {L.~P.}\ \bibnamefont
  {García-Pintos}}\ and\ \bibinfo {author} {\bibfnamefont {A.}~\bibnamefont
  {del Campo}},\ }\bibfield  {title} {\bibinfo {title} {Quantum speed limits
  under continuous quantum measurements},\ }\href
  {https://doi.org/10.1088/1367-2630/ab099e} {\bibfield  {journal} {\bibinfo
  {journal} {New Journal of Physics}\ }\textbf {\bibinfo {volume} {21}},\
  \bibinfo {pages} {033012} (\bibinfo {year} {2019})}\BibitemShut {NoStop}%
\bibitem [{\citenamefont {García-Pintos}\ and\ \citenamefont {del
  Campo}(2021)}]{Pintos20211}%
  \BibitemOpen
  \bibfield  {author} {\bibinfo {author} {\bibfnamefont {L.~P.}\ \bibnamefont
  {García-Pintos}}\ and\ \bibinfo {author} {\bibfnamefont {A.}~\bibnamefont
  {del Campo}},\ }\bibfield  {title} {\bibinfo {title} {Limits to perception by
  quantum monitoring with finite efficiency},\ }\bibfield  {journal} {\bibinfo
  {journal} {Entropy}\ }\textbf {\bibinfo {volume} {23}},\ \href
  {https://doi.org/10.3390/e23111527} {10.3390/e23111527} (\bibinfo {year}
  {2021})\BibitemShut {NoStop}%
\bibitem [{\citenamefont {Molina-Vilaplana}\ and\ \citenamefont {del
  Campo}(2018)}]{Molina2018}%
  \BibitemOpen
  \bibfield  {author} {\bibinfo {author} {\bibfnamefont {J.}~\bibnamefont
  {Molina-Vilaplana}}\ and\ \bibinfo {author} {\bibfnamefont {A.}~\bibnamefont
  {del Campo}},\ }\bibfield  {title} {\bibinfo {title} {Complexity functionals
  and complexity growth limits in continuous mera circuits},\ }\href
  {https://doi.org/10.1007/JHEP08(2018)012} {\bibfield  {journal} {\bibinfo
  {journal} {Journal of High Energy Physics}\ }\textbf {\bibinfo {volume}
  {2018}},\ \bibinfo {pages} {12} (\bibinfo {year} {2018})}\BibitemShut
  {NoStop}%
\bibitem [{\citenamefont {Margolus}(2021)}]{Margolus2021}%
  \BibitemOpen
  \bibfield  {author} {\bibinfo {author} {\bibfnamefont {N.}~\bibnamefont
  {Margolus}},\ }\bibfield  {title} {\bibinfo {title} {Counting distinct states
  in physical dynamics},\ }\href {https://doi.org/10.48550/arXiv.2111.00297}
  {\bibfield  {journal} {\bibinfo  {journal} {arXiv:2111.00297}\ } (\bibinfo
  {year} {2021})}\BibitemShut {NoStop}%
\bibitem [{\citenamefont {Mackel}\ \emph {et~al.}(2022)\citenamefont {Mackel},
  \citenamefont {Yang},\ and\ \citenamefont {del Campo}}]{Mackel2022}%
  \BibitemOpen
  \bibfield  {author} {\bibinfo {author} {\bibfnamefont {N.~E.}\ \bibnamefont
  {Mackel}}, \bibinfo {author} {\bibfnamefont {J.}~\bibnamefont {Yang}},\ and\
  \bibinfo {author} {\bibfnamefont {A.}~\bibnamefont {del Campo}},\ }\bibfield
  {title} {\bibinfo {title} {Quantum alchemy and universal orthogonality
  catastrophe in one-dimensional anyons},\ }\href
  {https://doi.org/10.48550/arXiv.2210.10776} {\bibfield  {journal} {\bibinfo
  {journal} {arXiv:2210.10776}\ } (\bibinfo {year} {2022})}\BibitemShut
  {NoStop}%
\bibitem [{\citenamefont {Garc\'{\i}a-Pintos}\ \emph
  {et~al.}(2022)\citenamefont {Garc\'{\i}a-Pintos}, \citenamefont {Nicholson},
  \citenamefont {Green}, \citenamefont {del Campo},\ and\ \citenamefont
  {Gorshkov}}]{Pintos2021}%
  \BibitemOpen
  \bibfield  {author} {\bibinfo {author} {\bibfnamefont {L.~P.}\ \bibnamefont
  {Garc\'{\i}a-Pintos}}, \bibinfo {author} {\bibfnamefont {S.~B.}\ \bibnamefont
  {Nicholson}}, \bibinfo {author} {\bibfnamefont {J.~R.}\ \bibnamefont
  {Green}}, \bibinfo {author} {\bibfnamefont {A.}~\bibnamefont {del Campo}},\
  and\ \bibinfo {author} {\bibfnamefont {A.~V.}\ \bibnamefont {Gorshkov}},\
  }\bibfield  {title} {\bibinfo {title} {Unifying {Q}uantum and {C}lassical
  {S}peed {L}imits on {O}bservables},\ }\href
  {https://doi.org/10.1103/PhysRevX.12.011038} {\bibfield  {journal} {\bibinfo
  {journal} {Physical Review X}\ }\textbf {\bibinfo {volume} {12}},\ \bibinfo
  {pages} {011038} (\bibinfo {year} {2022})}\BibitemShut {NoStop}%
\bibitem [{\citenamefont {Mohan}\ \emph {et~al.}(2022)\citenamefont {Mohan},
  \citenamefont {Das},\ and\ \citenamefont {Pati}}]{Mohan}%
  \BibitemOpen
  \bibfield  {author} {\bibinfo {author} {\bibfnamefont {B.}~\bibnamefont
  {Mohan}}, \bibinfo {author} {\bibfnamefont {S.}~\bibnamefont {Das}},\ and\
  \bibinfo {author} {\bibfnamefont {A.~K.}\ \bibnamefont {Pati}},\ }\bibfield
  {title} {\bibinfo {title} {Quantum speed limits for information and
  coherence},\ }\href {https://doi.org/10.1088/1367-2630/ac753c} {\bibfield
  {journal} {\bibinfo  {journal} {New Journal of Physics}\ }\textbf {\bibinfo
  {volume} {24}},\ \bibinfo {pages} {065003} (\bibinfo {year}
  {2022})}\BibitemShut {NoStop}%
\bibitem [{\citenamefont {Carabba}\ \emph {et~al.}(2022)\citenamefont
  {Carabba}, \citenamefont {H{\"{o}}rnedal},\ and\ \citenamefont
  {Campo}}]{Carabba2022}%
  \BibitemOpen
  \bibfield  {author} {\bibinfo {author} {\bibfnamefont {N.}~\bibnamefont
  {Carabba}}, \bibinfo {author} {\bibfnamefont {N.}~\bibnamefont
  {H{\"{o}}rnedal}},\ and\ \bibinfo {author} {\bibfnamefont {A.~d.}\
  \bibnamefont {Campo}},\ }\bibfield  {title} {\bibinfo {title} {Quantum speed
  limits on operator flows and correlation functions},\ }\href
  {https://doi.org/10.22331/q-2022-12-22-884} {\bibfield  {journal} {\bibinfo
  {journal} {{Quantum}}\ }\textbf {\bibinfo {volume} {6}},\ \bibinfo {pages}
  {884} (\bibinfo {year} {2022})}\BibitemShut {NoStop}%
\bibitem [{\citenamefont {H{\"{o}}rnedal}\ \emph {et~al.}(2023)\citenamefont
  {H{\"{o}}rnedal}, \citenamefont {Carabba}, \citenamefont {Takahashi},\ and\
  \citenamefont {del Campo}}]{Hornedal2023geometricoperator}%
  \BibitemOpen
  \bibfield  {author} {\bibinfo {author} {\bibfnamefont {N.}~\bibnamefont
  {H{\"{o}}rnedal}}, \bibinfo {author} {\bibfnamefont {N.}~\bibnamefont
  {Carabba}}, \bibinfo {author} {\bibfnamefont {K.}~\bibnamefont {Takahashi}},\
  and\ \bibinfo {author} {\bibfnamefont {A.}~\bibnamefont {del Campo}},\
  }\bibfield  {title} {\bibinfo {title} {Geometric {O}perator {Q}uantum {S}peed
  {L}imit, {W}egner {H}amiltonian {F}low and {O}perator {G}rowth},\ }\href
  {https://doi.org/10.22331/q-2023-07-11-1055} {\bibfield  {journal} {\bibinfo
  {journal} {{Quantum}}\ }\textbf {\bibinfo {volume} {7}},\ \bibinfo {pages}
  {1055} (\bibinfo {year} {2023})}\BibitemShut {NoStop}%
\bibitem [{\citenamefont {Shrimali}\ \emph {et~al.}(2022)\citenamefont
  {Shrimali}, \citenamefont {Bhowmick}, \citenamefont {Pandey},\ and\
  \citenamefont {Pati}}]{Pandey22}%
  \BibitemOpen
  \bibfield  {author} {\bibinfo {author} {\bibfnamefont {D.}~\bibnamefont
  {Shrimali}}, \bibinfo {author} {\bibfnamefont {S.}~\bibnamefont {Bhowmick}},
  \bibinfo {author} {\bibfnamefont {V.}~\bibnamefont {Pandey}},\ and\ \bibinfo
  {author} {\bibfnamefont {A.~K.}\ \bibnamefont {Pati}},\ }\bibfield  {title}
  {\bibinfo {title} {Capacity of entanglement for a nonlocal hamiltonian},\
  }\href {https://doi.org/10.1103/PhysRevA.106.042419} {\bibfield  {journal}
  {\bibinfo  {journal} {Phys. Rev. A}\ }\textbf {\bibinfo {volume} {106}},\
  \bibinfo {pages} {042419} (\bibinfo {year} {2022})}\BibitemShut {NoStop}%
\bibitem [{\citenamefont {Pandey}\ \emph
  {et~al.}(2023{\natexlab{a}})\citenamefont {Pandey}, \citenamefont {Shrimali},
  \citenamefont {Mohan}, \citenamefont {Das},\ and\ \citenamefont
  {Pati}}]{Pandey2022}%
  \BibitemOpen
  \bibfield  {author} {\bibinfo {author} {\bibfnamefont {V.}~\bibnamefont
  {Pandey}}, \bibinfo {author} {\bibfnamefont {D.}~\bibnamefont {Shrimali}},
  \bibinfo {author} {\bibfnamefont {B.}~\bibnamefont {Mohan}}, \bibinfo
  {author} {\bibfnamefont {S.}~\bibnamefont {Das}},\ and\ \bibinfo {author}
  {\bibfnamefont {A.~K.}\ \bibnamefont {Pati}},\ }\bibfield  {title} {\bibinfo
  {title} {Speed limits on correlations in bipartite quantum systems},\ }\href
  {https://doi.org/10.1103/PhysRevA.107.052419} {\bibfield  {journal} {\bibinfo
   {journal} {Physical Review A}\ }\textbf {\bibinfo {volume} {107}},\ \bibinfo
  {pages} {052419} (\bibinfo {year} {2023}{\natexlab{a}})}\BibitemShut
  {NoStop}%
\bibitem [{\citenamefont {Pandey}\ \emph
  {et~al.}(2023{\natexlab{b}})\citenamefont {Pandey}, \citenamefont {Bhowmick},
  \citenamefont {Mohan}, \citenamefont {Sohail},\ and\ \citenamefont
  {Sen}}]{Pandey2023BM}%
  \BibitemOpen
  \bibfield  {author} {\bibinfo {author} {\bibfnamefont {V.}~\bibnamefont
  {Pandey}}, \bibinfo {author} {\bibfnamefont {S.}~\bibnamefont {Bhowmick}},
  \bibinfo {author} {\bibfnamefont {B.}~\bibnamefont {Mohan}}, \bibinfo
  {author} {\bibnamefont {Sohail}},\ and\ \bibinfo {author} {\bibfnamefont
  {U.}~\bibnamefont {Sen}},\ }\bibfield  {title} {\bibinfo {title} {Fundamental
  speed limits on entanglement dynamics of bipartite quantum systems},\ }\href
  {https://doi.org/10.48550/arXiv.2303.07415} {\bibfield  {journal} {\bibinfo
  {journal} {arXiv:2303.07415}\ } (\bibinfo {year}
  {2023}{\natexlab{b}})}\BibitemShut {NoStop}%
\bibitem [{\citenamefont {Carmichael}(2007)}]{Carmichael2007}%
  \BibitemOpen
  \bibfield  {author} {\bibinfo {author} {\bibfnamefont {H.~J.}\ \bibnamefont
  {Carmichael}},\ }\href
  {https://doi.org/https://doi.org/10.1007/978-3-540-71320-3} {\emph {\bibinfo
  {title} {Statistical Methods in Quantum Optics 2: Non-Classical Fields}}}\
  (\bibinfo  {publisher} {Springer Science \& Business Media},\ \bibinfo {year}
  {2007})\BibitemShut {NoStop}%
\bibitem [{\citenamefont {Pati}\ \emph {et~al.}(2015)\citenamefont {Pati},
  \citenamefont {Singh},\ and\ \citenamefont {Sinha}}]{Pati2015}%
  \BibitemOpen
  \bibfield  {author} {\bibinfo {author} {\bibfnamefont {A.~K.}\ \bibnamefont
  {Pati}}, \bibinfo {author} {\bibfnamefont {U.}~\bibnamefont {Singh}},\ and\
  \bibinfo {author} {\bibfnamefont {U.}~\bibnamefont {Sinha}},\ }\bibfield
  {title} {\bibinfo {title} {Measuring non-hermitian operators via weak
  values},\ }\href {https://doi.org/10.1103/PhysRevA.92.052120} {\bibfield
  {journal} {\bibinfo  {journal} {Physical Review A}\ }\textbf {\bibinfo
  {volume} {92}},\ \bibinfo {pages} {052120} (\bibinfo {year}
  {2015})}\BibitemShut {NoStop}%
\bibitem [{\citenamefont {Hall}\ \emph {et~al.}(2016)\citenamefont {Hall},
  \citenamefont {Pati},\ and\ \citenamefont {Wu}}]{hall2016}%
  \BibitemOpen
  \bibfield  {author} {\bibinfo {author} {\bibfnamefont {M.~J.~W.}\
  \bibnamefont {Hall}}, \bibinfo {author} {\bibfnamefont {A.~K.}\ \bibnamefont
  {Pati}},\ and\ \bibinfo {author} {\bibfnamefont {J.}~\bibnamefont {Wu}},\
  }\bibfield  {title} {\bibinfo {title} {Products of weak values: Uncertainty
  relations, complementarity, and incompatibility},\ }\href
  {https://doi.org/10.1103/PhysRevA.93.052118} {\bibfield  {journal} {\bibinfo
  {journal} {Physical Review A}\ }\textbf {\bibinfo {volume} {93}},\ \bibinfo
  {pages} {052118} (\bibinfo {year} {2016})}\BibitemShut {NoStop}%
\bibitem [{\citenamefont {Mondal}\ \emph {et~al.}(2017)\citenamefont {Mondal},
  \citenamefont {Bagchi},\ and\ \citenamefont {Pati}}]{mondal2017}%
  \BibitemOpen
  \bibfield  {author} {\bibinfo {author} {\bibfnamefont {D.}~\bibnamefont
  {Mondal}}, \bibinfo {author} {\bibfnamefont {S.}~\bibnamefont {Bagchi}},\
  and\ \bibinfo {author} {\bibfnamefont {A.~K.}\ \bibnamefont {Pati}},\
  }\bibfield  {title} {\bibinfo {title} {Tighter uncertainty and reverse
  uncertainty relations},\ }\href {https://doi.org/10.1103/PhysRevA.95.052117}
  {\bibfield  {journal} {\bibinfo  {journal} {Physical Review A}\ }\textbf
  {\bibinfo {volume} {95}},\ \bibinfo {pages} {052117} (\bibinfo {year}
  {2017})}\BibitemShut {NoStop}%
\bibitem [{\citenamefont {Yu}\ \emph {et~al.}(2019)\citenamefont {Yu},
  \citenamefont {Jing},\ and\ \citenamefont {Li-Jost}}]{yu2019}%
  \BibitemOpen
  \bibfield  {author} {\bibinfo {author} {\bibfnamefont {B.}~\bibnamefont
  {Yu}}, \bibinfo {author} {\bibfnamefont {N.}~\bibnamefont {Jing}},\ and\
  \bibinfo {author} {\bibfnamefont {X.}~\bibnamefont {Li-Jost}},\ }\bibfield
  {title} {\bibinfo {title} {Strong unitary uncertainty relations},\ }\href
  {https://doi.org/10.1103/PhysRevA.100.022116} {\bibfield  {journal} {\bibinfo
   {journal} {Physical Review A}\ }\textbf {\bibinfo {volume} {100}},\ \bibinfo
  {pages} {022116} (\bibinfo {year} {2019})}\BibitemShut {NoStop}%
\end{thebibliography}%
\end{document}